\documentclass[twocolumn,traditabstract]{aa}
\usepackage{amssymb}
\usepackage{mathptmx}
\usepackage{natbib}
\bibpunct{(}{)}{;}{a}{}{,}

\usepackage[]{graphicx}
\usepackage{cool}
\usepackage{subcaption}
\usepackage{amsmath}
\usepackage{lscape}
\usepackage{txfonts}
\usepackage{hyperref}
\hypersetup{colorlinks=true,citecolor=blue}
\usepackage{units}
\usepackage{csquotes}

\usepackage{comment}
\usepackage{verbatim}
\usepackage{url}
\usepackage[T1]{fontenc}

\usepackage{xcolor}     % Text color
\usepackage{soul}       % for strikeout, can be removed after comments are treated
\usepackage{array}
\usepackage{xspace}
\usepackage{fontawesome}

\begin{document}

%\defcitealias{Han22}{W22}

\title{Strong-lensing and kinematic analysis of CASSOWARY 31: Can strong lensing constrain the masses of multi-plane lenses? }

\author{H.~Wang\inst{1,2} \and R.~Ca\~nameras\inst{1,2,3} \and S.~H.~Suyu\inst{1,2,4}  \and A.~Galan\inst{2,1} \and C.~Grillo\inst{5,6}
\and G.~B.~Caminha\inst{2,1} \and L.~Christensen\inst{7}}

\institute{
Max-Planck-Institut f\"ur Astrophysik, Karl-Schwarzschild-Str. 1, 85748 Garching, Germany \\
{\tt e-mail: wanghan@mpa-garching.mpg.de}
\and
Technical University of Munich, TUM School of Natural Sciences, Physics Department, James-Franck Str. 1, 85748 Garching, Germany
\and
Aix Marseille University, CNRS, CNES, LAM, Marseille, France
\and
Academia Sinica Institute of Astronomy and Astrophysics (ASIAA), 11F of ASMAB, No. 1, Section 4, Roosevelt Road, Taipei 10617, Taiwan
\and
Dipartimento di Fisica, Universit\`a degli Studi di Milano, via Celoria 16, I-20133 Milano, Italy
\and
INAF-IASF Milano, via A. Corti 12, I-20133 Milano, Italy
\and
Cosmic Dawn Center, Niels Bohr Institute, Univ. of Copenhagen, Jagtvej 128, 2200 Copenhagen, Denmark
}

\titlerunning{Can strong lensing constrain the masses of multi-plane lenses}

\authorrunning{H.Wang et al.} \date{Received / Accepted}

\abstract{We present a mass measurement for the secondary lens along the line of sight (LoS) from the multi-plane strong lens modeling of the group-scale lens CASSOWARY 31 (CSWA\,31). The secondary lens at redshift $z = 1.49$ is a spiral galaxy well aligned along the LoS with the main lens at $z = 0.683$. Using the MUSE integral-field spectroscopy of this spiral galaxy, we measured its rotation velocities and determined the mass from the gas kinematics. We compared the mass estimation of the secondary lens from the lensing models to the mass measurement from kinematics, finding that the predictions from strong lensing  tend to be higher. By introducing an additional lens plane at $z = 1.36$ for an overdensity known to be present, we find a mass of $\simeq 10^{10}$ M$_\odot$ enclosed within 3.3 kpc of the centroid of the spiral galaxy, which approaches the estimate from kinematics. This shows that secondary-lens mass measurements from multiple-plane modeling are affected by systematic uncertainties from the degeneracies between lens planes and the complex LoS structure. Conducting a detailed analysis of the LoS structures is therefore essential to improve the mass measurement of the secondary lens.}


\maketitle
\section{Introduction}
\label{sec:intro}
In the $\rm  \Lambda$ cold dark matter model, galaxies grow hierarchically via the accretion of cold gas from the circumgalactic medium and mergers with nearby galaxies. Measuring robust galaxy masses over broad redshift ranges is key to improving our understanding of this broad picture \citep[e.g.,][]{2010ApJ...721..193P}. Yet, at high redshifts ($z \geq 1$), direct measurements are challenging due to observational limitations, and mass estimates not only rely on assumptions derived from our knowledge of the nearby Universe, but are also prone to several systematics. For instance, for gas-rich spiral galaxies, dynamical masses inferred from gas kinematics can be significantly biased in the case of departures from smooth rotation due to secular instabilities \citep[e.g.,][]{secularinst}, merging events, or feedback processes \citep[e.g.,][]{shockoutflow}.

In this context, strong lensing has been established as a robust, accurate tracer of the total mass and inner structure of foreground galaxies \citep[e.g.,][]{koopmans06, auger10, Anowar}, galaxy groups, and clusters \citep[e.g.,][]{Tommaso, Meneghetti}. Photons traveling from distant sources get deflected by an overdense region associated with the foreground galaxies and form arcs or a full Einstein ring around the deflectors. The locations of the lensed images and the surface brightness of the extended arcs can be used to constrain the mass distribution of the foreground deflectors. 

In most strong lensing scenarios, a single overdense region along the path between the source and observer is considered for modeling. The line-of-sight (LoS) structure is neglected, which is usually a sensible approximation since the main lens contributes dominantly to the deflection angle \citep[e.g.,][]{doubellens}. In some cases, galaxies aligned with the main lens can nonetheless produce non-negligible perturbations, particularly for group- and cluster-scale lenses, which tend to have rich LoS structures \citep{Bayliss14}. To reconstruct the lensed sources and improve the root mean square (rms) of the predicted image position for such systems, a secondary lens needs to be included when computing multiple-lens-plane models \citep[e.g.,][]{Schneidermulti,Wong+17}.

In this work we further explored the constraints on the mass of secondary lenses inferred from multi-plane lens models by conducting a case study based on the group-scale lens CASSOWARY 31 \citep[CSWA 31;][]{belokurov09,stark13}. In \citet{Han22}, through detailed modeling of the multi-scale mass distribution, we found that this system comprises a spiral galaxy at $z = 1.49$ that is well aligned with the main lens at $z = 0.683$, with both structures contributing significantly to the deflection angles of source galaxies at higher redshifts. In addition to strong lensing, gas kinematics inferred from integral-field spectroscopy (IFS) with the Multi-Unit Spectroscopic Explorer (MUSE; \citealt{muse}) allowed us to infer an independent mass estimate for this secondary lens. While such known systems with two or more deflectors are rare, their number is expected to increase with the next generation of deep, wide-field imaging surveys such as the Legacy Survey of Space and Time, \textit{Euclid}, and \textit{Roman} \citep[e.g.,][]{Collett+2015, mandelbaum18}. It is therefore becoming essential to investigate multi-plane lens modeling as an alternative mass probe to study the inner structure of galaxies acting as secondary deflectors. 

\begin{figure*}
    \centering
    \begin{subfigure}[b]{0.63\textwidth}
        \includegraphics[width=\textwidth]{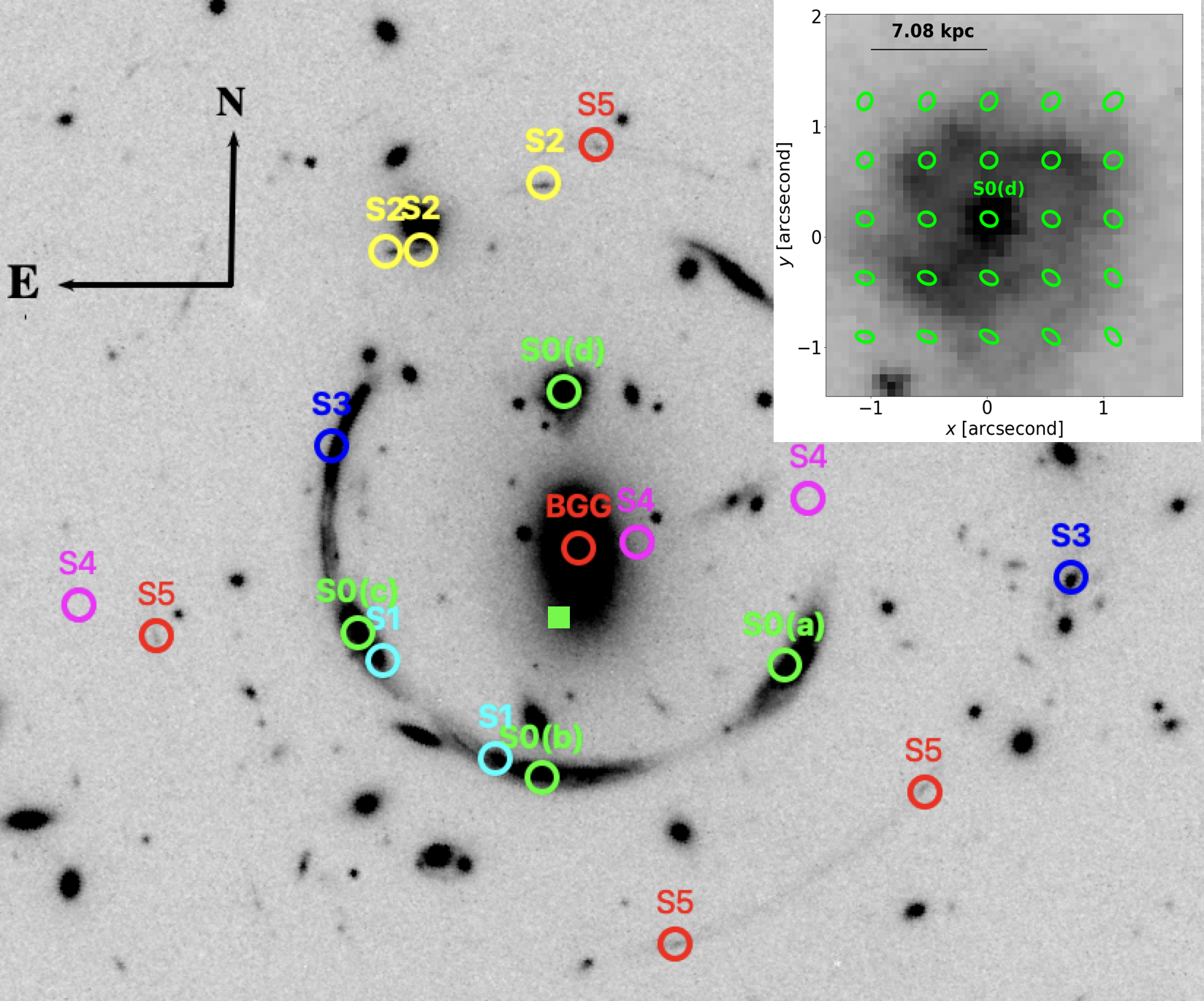}
    \end{subfigure}
    \hfill
    \begin{subfigure}[b]{0.36\textwidth}
        \includegraphics[width=\textwidth]{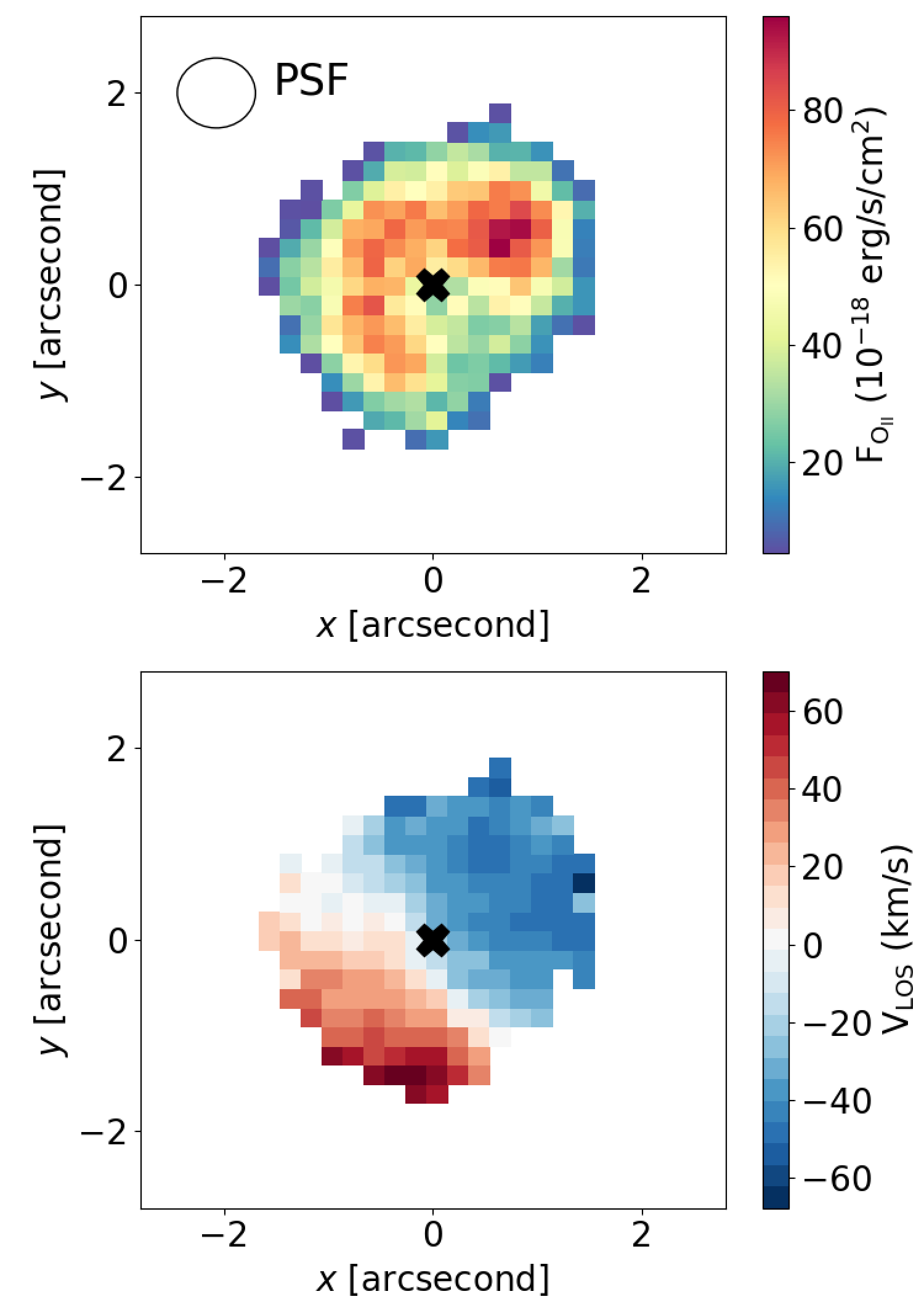}
    \end{subfigure}
    \caption{HST imaging and kinematics of the lensed galaxy S0.  \textit{Left panel:}  HST/WFC3 image of CSWA 31 in the F160W band. The orientation is marked by the black arrows (north upward and east leftward) and kept fixed for other panels. The BGG at $z = 0.683$ is surrounded by six  multiple-image sets at different redshifts marked with circles, using a different color for each set. The image sets S0 and S1 are bright features from the same spiral galaxy at $z = 1.49$, with a zoomed-in image of S0(d) in the right inset. The green square shows the de-lensed S0 position in the HST imaging. The green ellipses in the zoomed-in panel illustrate the minor variations in distortion over S0(d) caused by strong lensing. \textit{Right panel:}  [OII] intensity map (top) and the LoS velocity map (bottom) for the S0(d) image extracted from MUSE IFS. The FWHM of the MUSE PSF is shown in the upper left in the intensity map. The bold black cross shows the adopted S0(d) centroid determined from $^{\tt 3D}${\tt Barolo}.}
    \label{fig:combine_int_vel}
\end{figure*}

The paper is organized as follows. In Sect.~\ref{sec:Data} we briefly introduce the properties of CSWA\,31 and the datasets. In Sect.~\ref{sec:method} we present the independent mass measurements from strong lensing and kinematic modeling. In Sect.~\ref{sec:discu} we discuss the uncertainties of each method and compare the results. We also present an updated model with an additional lens plane. In Sect.~\ref{sec:summary} we summarize our findings and discuss the validity of mass predictions for secondary lenses obtained from multi-plane lens modeling. Throughout this work we assume $H_{\rm 0} = 70$\,km s$^{-1}$ Mpc$^{-1}$, $\Omega_{\rm m} = 0.3,$ and $\Omega_{\Lambda} = 0.7$. Hence, 1\arcsec\ corresponds to 7.08\,kpc at the main lens redshift of $z = 0.683$, and to 8.46\,kpc at the secondary lens redshift of $z = 1.49$.

\section{Data}
\label{sec:Data}
CSWA~31 is a group-scale lens with a central ultra-massive brightest group galaxy (BGG) at $z = 0.683$, surrounded by six sets of multiple images from five independent background sources. The high-resolution \textit{Hubble} Space Telescope (HST) imaging in F160W and F438W filters, the VLT/MUSE spectroscopy of the field (program 0104.A-0830(A), PI: Ca\~nameras), and the joint analysis of these data are described in detail in \citet{Han22}. The background spiral galaxy at $z = 1.49$ studied in this letter and referred to as S0, is lensed into four multiple images forming an extended arc (see Fig.~\ref{fig:combine_int_vel}). We identified five additional sets of multiple images -- S1, S2, S3, S4, and S5 -- at four independent redshifts using MUSE spectroscopic and HST (F160W$-$F438W) colors. Image sets S1 and S2 are at redshift $z = 1.49$ and S3, S4 and S5 at redshifts $z = 2.76,~3.43,$ and $z = 4.21,$ respectively. In addition, 46 group members located within $\pm 3000$~km s$^{-1}$ of the BGG redshift were selected from the MUSE spectroscopic catalog.

The spiral galaxy S0 displays a disk-like morphology in HST/F160W, with a bright bulge and three main spiral arms, and is apparently oriented nearly face-on. The best-fit Sérsic profile of image S0(d) shows that this galaxy has an effective radius $R_{\rm eff}$ of 1.17\arcsec, or about 3.8~kpc in the source plane (after correcting for magnification), and an axis ratio $q = 0.85$. The properties of S0 are similar to the strongly lensed spiral galaxy at $z \approx 1.5$ hosting SN Refsdal, for which \citet{diteodoro18} inferred 3D kinematic models.

In this work, we focus on the least distorted image S0(d), shown in Fig.~\ref{fig:combine_int_vel} (left inset), to conduct a kinematic analysis of S0 in the image plane \citep[see also][]{diteodoro18}. We verify that spatial variations of lensing shear are negligible for S0(d) using the mass model Esr2$-$$\text{MP}_{\rm test}$ from \citet{Han22} ({\tt baseline} model hereafter). We use this model to map small circles of 0.052\arcsec\ radius distributed over the source plane at $z = 1.49$, to the image S0(d). The circular sources get magnified and distorted into ellipses in the image plane, but these ellipses do not vary significantly in size and orientation over S0(d), suggesting overall uniform lensing distortions over this image (see the green ellipses on the inset of Fig.~\ref{fig:combine_int_vel} left panel). In addition, the magnification factor $\mu \simeq 7$ of image S0(d) offers a good opportunity to zoom in onto the gas kinematics of this spiral.

The MUSE data cube has a point spread function (PSF) with a full width at half maximum (FWHM) of $0.75\arcsec \times 0.73$\arcsec. Its wavelength resolution is $\text{FWHM} = 2.35~\AA$, with a channel width of $1.25~\AA$. To focus on the [OII] doublet, the cube was cropped between $3719~\AA$ to $3734~\AA$ in the rest frame of S0. The [OII] line intensity map of S0(d) shows an asymmetrical feature, indicating that ionized gas reservoirs form denser regions along the spiral arms seen in stellar continuum with HST (see Fig.~\ref{fig:combine_int_vel}). The total [OII] flux over the source is $203.8 \times 10^{-18}$\,erg\,s$^{-1}$. The LoS velocity map is derived by fitting double Gaussian functions to each pixel, representing a preliminary step to investigate the presence of velocity gradients within the S0(d) image. Additionally, these maps offer initial values for geometrical parameters in $^{\tt 3D}${\tt Barolo}.

\section{Methodology}
\label{sec:method}
To investigate the constraints on secondary lens masses, we inferred two independent mass estimates of the S0 spiral. In Sect.~\ref{sec:dyn_mass} we present the kinematic modeling procedure and the mass measurement from the best-fit rotation curve. In Sect.~\ref{sec:lens_mod} we present the multi-plane strong lens modeling and the S0 mass measurement from the lensing approach.

\subsection{Mass measurement from kinematic modeling}
\label{sec:dyn_mass}
Spiral galaxies at $z \geq 1$ are thought to settle into marginally stable, rotating disks with turbulence \citep[e.g.,][]{turbulentdisk2009, RC100}. The apparent velocity gradient in Fig.~\ref{fig:combine_int_vel} and the smooth disk-like morphology from HST suggest a rotating velocity field. We adopted $^{\tt 3D}${\tt Barolo} \citep{3Dbarolo} to derive a simple rotating disk model of S0 using the MUSE data of image S0(d), following a similar approach as \citet{diteodoro18}. This image is well de-blended from other galaxies in the field and the minor distortions of S0(d) caused by strong lensing vary on scales smaller than the MUSE PSF. The beam-smearing of galaxy kinematics, which is accounted for by $^{\tt 3D}${\tt Barolo}, dominates over systematic errors due to lensing shear.

$^{\tt 3D}${\tt Barolo} initially identifies a mask region within the MUSE data cube encompassing the emission lines. Subsequently, it builds a kinematic model convolved with the MUSE PSF to characterize these lines by using tilted rings. In our case, $^{\tt 3D}${\tt Barolo} replicates the [OII] emission doublet at $\lambda = 3726~\AA$ and $3729~\AA$. Each ring is defined with a centroid, position angle, inclination angle $i$, systematic velocity $V_{\rm sys}$, rotation velocity $V_{\text{rot}}$ and velocity dispersion $\sigma_{\text{v}}$. We built the kinematic model by using four tilted rings, with a 0.4\arcsec\ width, and adopting the S0(d) centroid in HST imaging with the S0 position angle determined from the velocity gradient (see Fig.~\ref{fig:combine_int_vel}) as initial values. We fixed the $V_{\rm sys}$ in all tilted rings to 0 km s$^{-1}$. We iteratively refined these values with $^{\tt 3D}${\tt Barolo} and subsequently allowed variations in only the $V_{\text{rot}}$ and $\sigma_{\text{v}}$ at each ring (see Fig.~\ref{fig:pv_major}). In the fitting, we adopted the {\tt local} normalization and a uniform weighting function\footnote[1]{For the “type-two” residual defined in Eqs.~(2) and 3(b) in Sect.~2.4 of \citet{3Dbarolo}.} to account for the asymmetry in the intensity map of S0(d).
 
\begin{figure}
  \centering
  \includegraphics[width=1.\linewidth]{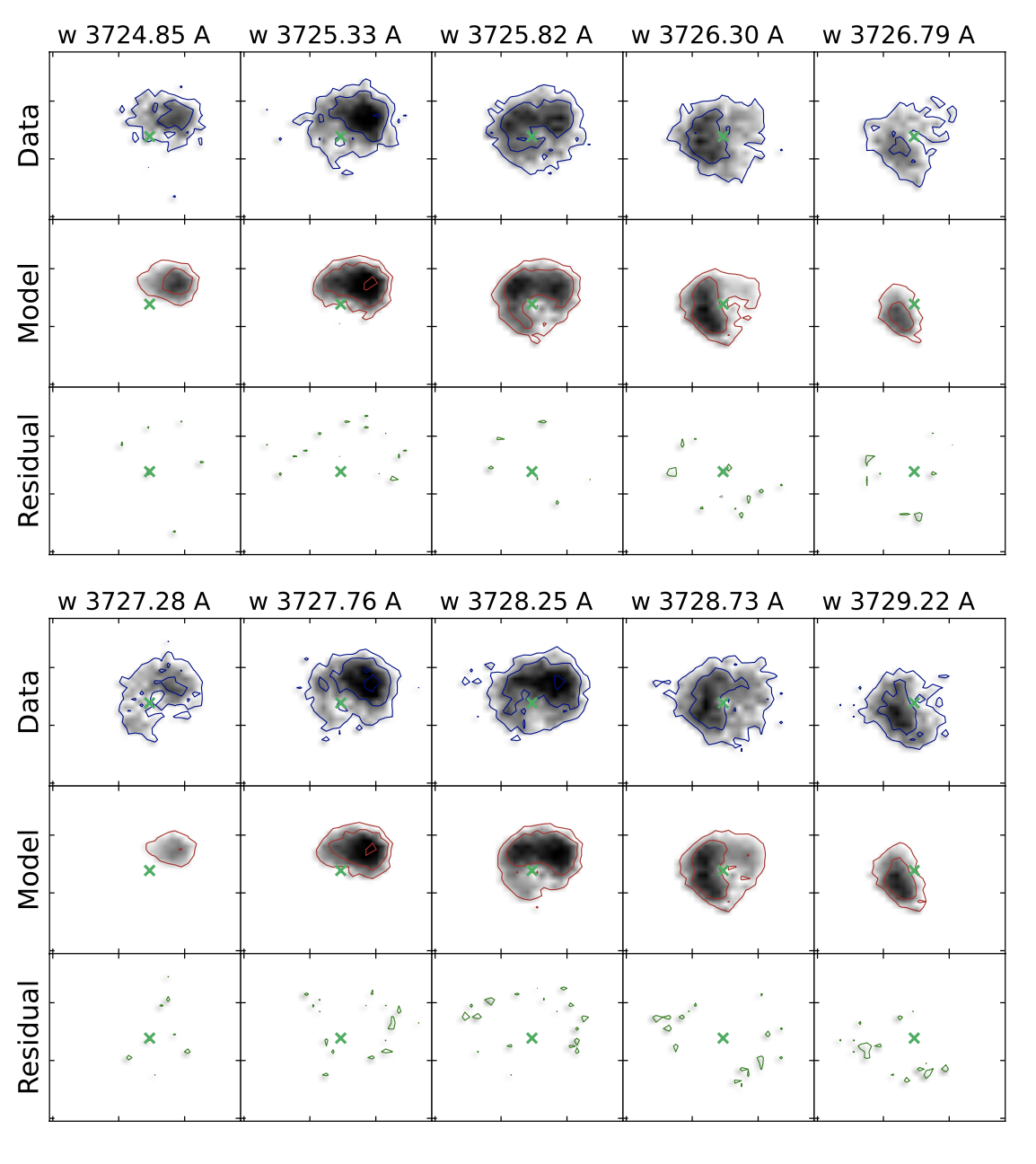}
  \caption{Comparison between MUSE IFS of image S0(d) (blue contours) and the $^{\tt 3D}${\tt Barolo} model (red contours). The third row in the upper and bottom panels shows the residuals. Contour levels are set at $2\sigma n$, where the $\sigma$ noise $=2 \times 10^{-20}$~erg s$^{-1}$~cm$^{-2}$ $\AA^{-1}$ and $n=1$, 2, 4, 8. The green cross shows the centroid position of S0(d). This plot shows the fitting result for the two lines of the [OII] emission doublet at (rest frame) $\lambda = 3726~\AA$ and $3729~\AA$ in the upper and bottom panels, respectively. 
  }
  \label{fig:channel_map.pdf}
\end{figure}

$^{\tt 3D}${\tt Barolo} projects the tilted ring model to the plane of the sky using $i$ and compares it to the observational data. The inclination angle is an important parameter, but it is hard to estimate in practice due to the unknown intrinsic shape of galaxies. Without losing any generality, we built a range of kinematic models covering $i$ from $5^\circ$ to $35^\circ$ (see the discussion in Sect.~\ref{subsec:syst_kin}). The inclination is not well constrained, and Fig.~\ref{fig:channel_map.pdf} shows a typical best-fit model per wavelength channel, for $i=20.8^\circ$.

Due to the moderate spectral resolution of the MUSE data cube, velocity dispersion values returned by $^{\tt 3D}${\tt Barolo} decrease to $0$ km s$^{-1}$ for the inner radius of $0.2\arcsec$ and the outer radius of $1.4\arcsec$, for all kinematic models with inclination angles between $5^\circ$ and $35^\circ$. To address this limitation, we reduced the number of rings to three and obtained the best-fit $\sigma_{\rm v}$ at distances of $0.6\arcsec$, $0.9\arcsec$, and $1.2\arcsec$ from the centroid of image S0(d), finding an average value of $\sigma_{\rm v} = 41.3$ km s$^{-1}$. However, the residuals in the channel maps are worse than the model built with four rings (see Fig.~\ref{fig:channel_map_three_rings}). We thus decided to retain the model with four rings and solely adopted the $V_{\text{rot}}$ along the major axis at the second ring at $r = 0.6\arcsec$ and the third ring at $r = 1.0\arcsec$ to determine the dynamical mass of S0 (see Fig.~\ref{fig:rotation_curve}).  

We determined the total mass of S0 using $V_{\text{rot}}$ via\begin{equation}
     M_\text{rot} (r') = \frac{V_\text{rot}^2}{G}r',
     \label{eq:rot mass}
\end{equation}
where $r'$ is the radial distance from the S0 centroid in the source plane at $z = 1.49$. Strong lensing magnifies a circular source with area $\pi {r'}^2$ by a factor of $\mu$ in the image plane (see Table~\ref{tab:theta_E_s0}). We converted the radii $r$ in the S0(d) image to radii $r'$ in the source plane via
\begin{equation}
     r' = \frac{r}{\sqrt{\mu}}
     \label{eq:r_convert}
.\end{equation}
We obtained a total mass of S0 of $\simeq 10^{10}$~M$_{\odot}$ at 3.3~kpc in the source plane, which is about the effective radius of S0 at $z=1.49$ (see Fig.~\ref{fig:mass_s0}).

\subsection{Mass measurement from multi-plane lens modeling}
\label{sec:lens_mod}
We summarize in Table~\ref{tab:theta_E_s0} the different lens models we considered in this work. In this section, we adopt the multi-plane lens models Img$-$2MP, Img$-$2MP(L/D) and {\tt baseline} from \citet{Han22}. The average magnification factor, $\mu,$ of S0(d) has minor variations among these models, and the induced shear is negligible, as shown in Fig.~\ref{fig:combine_int_vel} for the {\tt baseline} model.

These three models consist of two lens planes, and were obtained with the {\tt GLEE} software \citep{suyu2010,Suyu2012}. The main lens plane at $z = 0.683$ incorporates the mass contributions from the BGG, 46 group members confirmed by MUSE spectroscopy, the group halo, and a constant external shear to account for the dense environment surrounding CSWA\,31. The mass of the BGG and group members are modeled with truncated dual pseudo-isothermal elliptical \citep[dPIE;][]{dpie2007,dpie2010} mass profiles. Due to the limited number of constraints from lensing, the Einstein radii and truncation radii of group members are scaled with their luminosities in the $F160W$ band following the approach in \cite{Grillo2015} and \cite{Chirivi2018}. The underlying group halo is depicted by a softened power-law elliptical mass profile with an extended core radius \citep{spemd}. The secondary lens plane at $z = 1.49$ accounts for the mass of the spiral galaxy S0, which serves as a lensed source, being deflected by the main lens. Additionally, as a secondary lens along the LoS, S0 deflects the light from the background sources S3, S4, and S5 at higher redshifts. Accounting for this secondary strong deflector (S0) improves the fitting of the image positions of S3, S4, and S5, while also enabling the simultaneous reconstruction of the extended sources S0 and S3 in the {\tt baseline} model. Excluding the secondary lens plane in the modeling results in the failure to reconstruct S3 and leads to higher errors in the predicted image positions of S3, S4, and S5.

In all three models, S0 is modeled with a spherical isothermal (SIS) profile, which has a 3D density profile described as
\begin{equation}
    \rho(r) = \frac{{\sigma_{\text v}}^{2}}{2\pi G r^2},
    \label{eq: 3d density}
\end{equation}
where $\sigma_{\text v}$ is the velocity dispersion, related to the Einstein radius $\theta_{\rm E}$ (for a lensed background source at infinite redshift) via ${\sigma_{\text v}}^{2} = \frac{\theta_{\text E} c^2}{4 \pi}$. Thus, Eq.~(\ref{eq: 3d density}) can be written in terms of $\theta_{\text E}$:
\begin{equation}
    \rho(r) = \frac{{\theta_{\text E} c^2}}{8 {\pi}^{2} G r^2}.
\end{equation}
We calculated the mass of S0 enclosed within radius $r'$ in spherical coordinates,
\begin{equation}
     M_{\text {lens}} (r') = \int_{0}^{r'} 4\pi r^2 \frac{{\theta_{\text E} c^2}}{8 {\pi}^{2} G r^2 }\, d{r}
,\end{equation}
such that
\begin{equation}
    M_{\text {lens}} (r') =   \frac{{\theta_{\text E} c^2}}{2 {\pi}G} r',
    \label{eq:mass lens}
\end{equation}
where ${r'}$ is the radial distance in the source plane.  Eq.~(\ref{eq:mass lens}) is valid in the inner region of the galaxy within the halo truncation.

The lens models Img$-$2MP, Img$-$2MP(L/D), and {\tt baseline} adopted in this study rely on different sets of observational constraints to ascertain the best-fit values of the Einstein radius $\theta_{\rm E,S0}$ (see Table~\ref{tab:theta_E_s0}). The model Img$-$2MP adopts image positions of six multiple-image sets as constraints. In model Img$-$2MP(L/D), \citet{Han22} combines the stellar kinematics of the BGG with the multiple image positions, to perform a joint strong lensing and dynamical modeling. The {\tt baseline} model uses extended arcs of S0 and S3 in addition to the image positions of three multiple-image sets S2, S4, and S5 as constraints. In Fig.~\ref{fig:mass_s0} we plot the S0 mass distribution as a function of the source plane radius for these three models. The S0 mass predicted from the {\tt baseline} model is three times smaller than that for the other two, and we further analyze this discrepancy in Sect.~\ref{subsec:syst_lens}.

\begin{table}
  \caption{Strong lens models used in this study, their different constraints, the best-fit values of $\theta_{\rm E,S0}$, and the average magnification factors, $\mu,$ weighted by the surface brightness over the S0(d) image.}
\centering
\begin{tabular}{lccc}
\hline
\hline \\[-0.3em]
Model name & Constraints & $\theta_{\rm E,S0}$ &$\mu$\\ %[+0.5em]
\hline \\[-0.3em]
Img$-$2MP &  6 lensed-image sets  &  4.4\arcsec\ & 5.5 \\
Img$-$2MP (L/D)  &  6 sets + kinematics of BGG & 4.6\arcsec\ & 6.9 \\
{\tt baseline} & 3 sets + extended arcs of S0, S3 & 1.65\arcsec\ & 6.7 \\
Esr2$-$3MP & 3 sets + extended arcs of S0, S3 & 0.85\arcsec\ & 7.3 \\   
\hline
\end{tabular}
\tablefoot{The first three models share the same configuration and mass parametrization. In the last model Esr2-3MP, we accounted for the additional perturbation caused by foreground galaxies at $z =1.36$ and performed lens modeling with three lens planes.}
\label{tab:theta_E_s0}
\end{table}

\section{Mass comparison and discussion}
\label{sec:discu}
Strong lens models presented in Sect.~\ref{sec:lens_mod} tend to assign significantly more mass to S0 than the modeling of gas kinematics. We discuss this discrepancy by highlighting the main uncertainties from each approach in Sects.~\ref{subsec:syst_kin} and \ref{subsec:syst_lens}. In Sect.~\ref{subsub:Two lens plane modeling} we discuss the results from the two-plane lens models that are presented in \citet{Han22}. Then, in Sect.~\ref{subsub:Three lens planes modeling}, we present an additional model with a third lens plane to further explore the impact of LoS structures.

\subsection{Uncertainties from the kinematic modeling}
\label{subsec:syst_kin}
\begin{figure}
    \centering
    \begin{subfigure}[b]{0.5\textwidth}
        \includegraphics[width=\textwidth]{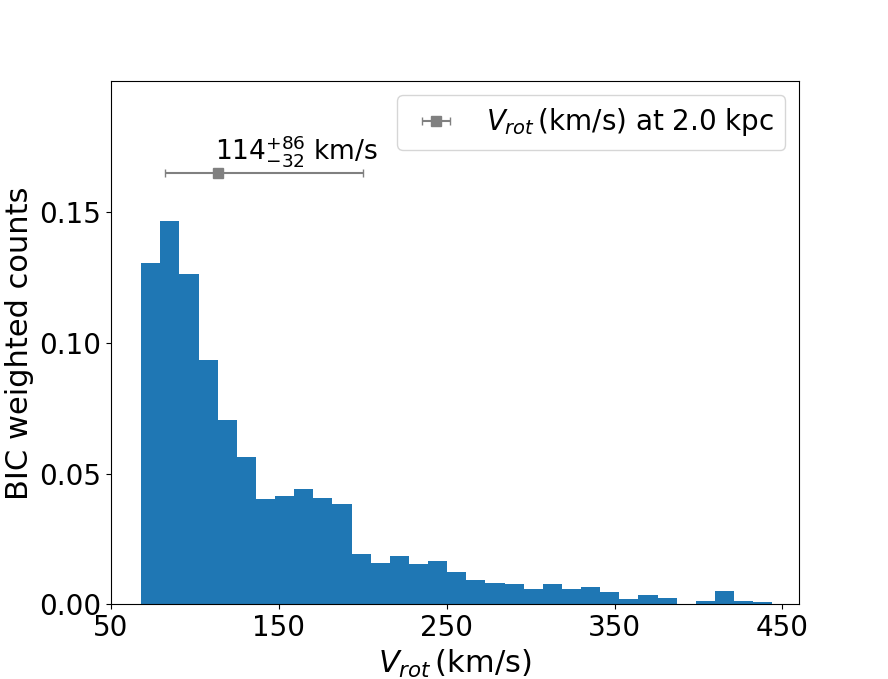}
    \end{subfigure}
    \hfill
    \begin{subfigure}[b]{0.5\textwidth}
        \includegraphics[width=\textwidth]{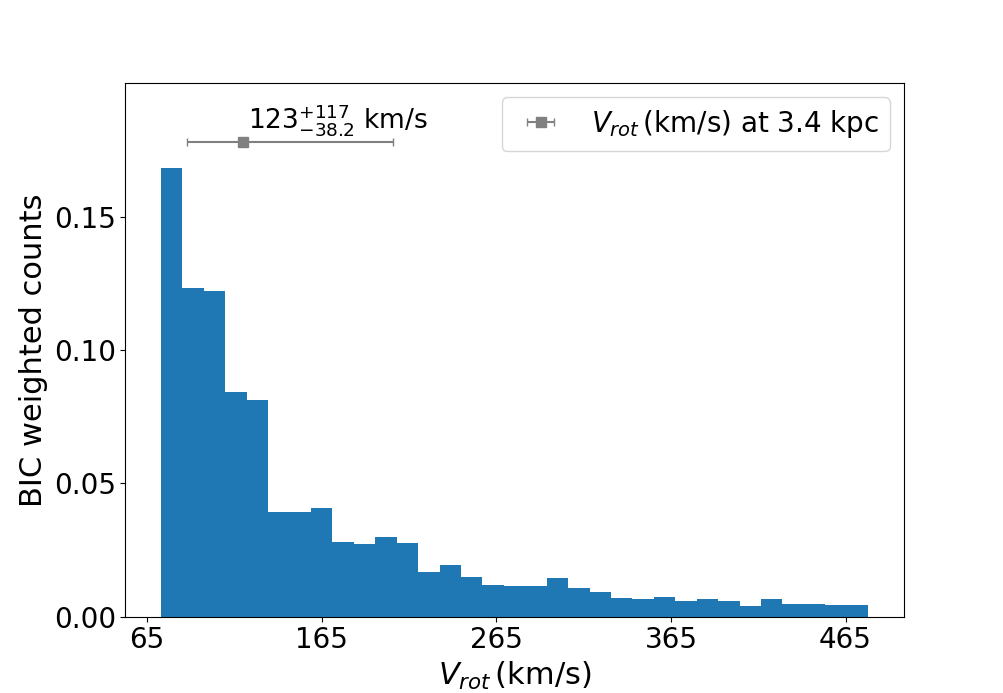}
    \end{subfigure}
    \caption{Histogram of $V_{\rm rot}$ weighted by the BIC. The upper and bottom panels show the weighted $V_{\rm rot}$ distribution at 2.0 kpc and at 3.3 kpc from the galaxy centroid in the source plane, respectively. The 1$\sigma$ uncertainties on $V_{\rm rot}$ are determined by drawing 5000 samples of the inclination angles in the range [5$^\circ$, 35$^\circ$]. }
    \label{fig:rotation_curve}
\end{figure}

The unknown inclination angle $i$ of the spiral has the largest contribution to the error budget on the mass inferred from kinematic modeling. Since S0 is observed nearly face-on, a small variation of $i$ leads to significant changes in the rotation velocity along the major axis, as $V_{\text{rot}} = V_{\text{LoS}}/ \sin{i}$. To obtain a sensible estimation of the uncertainty caused by $i$, we randomly drew 5000 samples from a broad range of angles $i$ over [5$^\circ$, 35$^\circ$] and fit the MUSE data cube for each sample. We performed a model comparison by making use of the Bayesian information criterion (BIC). Since we used the same data cube and only allowed the $V_{\rm rot}$ and $\sigma_{\rm v}$ to vary in each sample, the BIC in our case is computed by the log-likelihood term. We followed the approach of \citet{Birrer19} 
defining the BIC weight $w_{i}$ with respect to the minimal BIC value ${\rm BIC}_{\rm min}$ in our sample,
\begin{equation}
    w_{i}  = \frac{\exp{ \{ -  ({\rm BIC}_{i} - {\rm BIC}_{\rm min})/2 \} }}{\sum_{i=1}^{5000} \exp{\{-({\rm BIC}_{i} - {\rm BIC}_{\rm min})/2\}}}.
\end{equation}
We then determined the median $V_{\rm rot}$ and the 1$\sigma$ uncertainty from the 16th, 50th, and 84th of the weighted $V_{\rm rot}$ at radii of 0.6~\arcsec\ and 1.0~\arcsec\ (on the image plane) from the centroid of image S0(d), which correspond to 2.0\,kpc and 3.3\,kpc  in the source plane (see Fig.~\ref{fig:rotation_curve}). 

We obtained the median values of $V_{\rm rot} = 123$~km s$^{-1}$ and $\sigma_{\rm v} = 40$~km s$^{-1}$ at 3.3~kpc from the galaxy center, and a ratio $V_{\rm rot}/\sigma_{\rm v}$ of 3.1 showing that this spiral galaxy is indeed rotationally-supported. Median values are broadly consistent with the evolution of ionized gas kinematics of star-forming galaxies at $z \approx 1$--2 in the literature \citep[e.g.,][]{Wisnioski15, Ubler19, Genzel2020, price21, Puglisispiral}. With a similar modeling of the [OII] doublet, \citet{diteodoro18} characterized the kinematics of the spiral hosting SN Refsdal up to 8\,kpc from the galaxy center, probing into the flat portion of the rotation curve. While their similar $V_{\rm rot}$ and lower $\sigma_{\rm v} \simeq 27$~km s$^{-1}$ indicate stronger rotation support, the properties of both spirals at $z \approx 1.5$ are broadly consistent.

We used the value of $V_{\rm rot}$ at 3.3~kpc to place S0 on the stellar-mass Tully-Fisher relation (TFR). We obtained an estimate of the galaxy stellar mass by fitting the $g$-, $r$-, $i$-, and $z$-band photometry of image S0(d) from the DESI Legacy Surveys DR10, together with the F160W band from HST with CIGALE \citep{boquien19} and following the approach described in \citet{Han22}. This yields $M_{*} = (2.1 \pm 0.4) \times 10^{10}$ M$_\odot$, corrected by $\mu = 6.7$ from the {\tt baseline} model, showing that S0 follows the stellar-mass TFR at $z \simeq 0$ from \citet{Tullyfisher} (see Fig.~\ref{fig:TFR}). The stellar mass is also comparable to the total mass at $R_{\rm eff}$ from the kinematic analysis, showing that the baryonic component dominates the inner mass budget.

\begin{figure}
  \includegraphics[width=1.\linewidth]{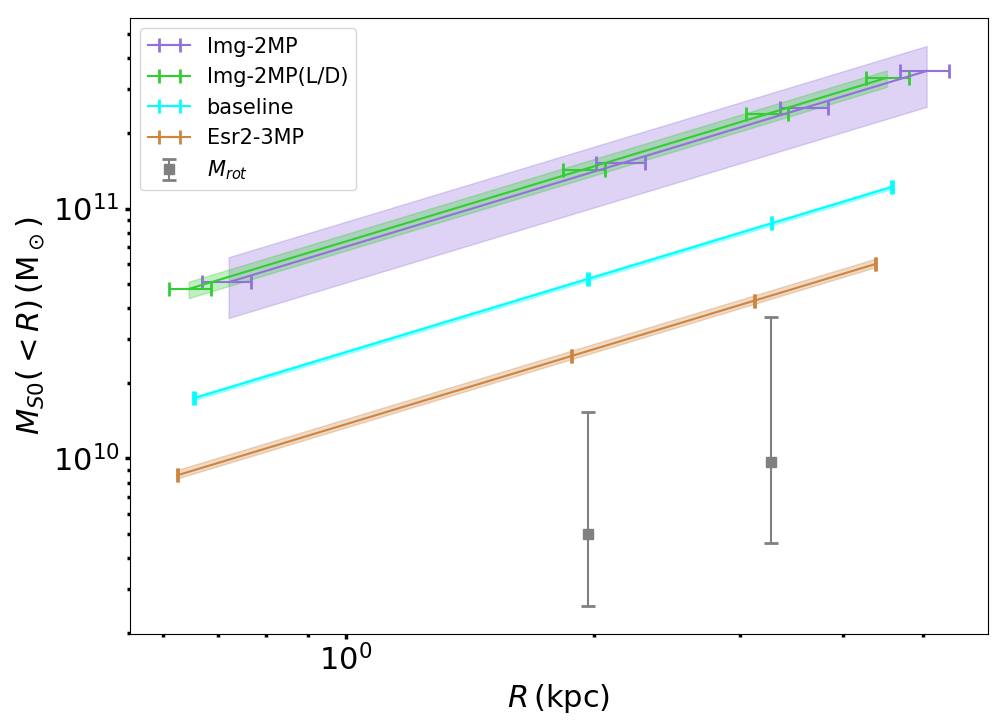}
  \caption{Comparison of mass distributions for the spiral galaxy S0 at source plane from strong lensing and kinematic modeling. The solid lines show the S0 mass estimated from lens models using Eq.~(\ref{eq:mass lens}), given different datasets as constraints (see Table~\ref{tab:theta_E_s0}). The statistical 1$\sigma$ uncertainty of the S0 mass from lens models is estimated from the 16th, 50th, and  84th percentiles of the posterior probability distribution functions (PDFs). The 1$\sigma$ uncertainty along the $x$-axis of the S0 mass is determined from the magnification modification using Eq.~(\ref{eq:r_convert}). The gray squares show the S0 mass estimated from kinematic modeling using Eq.~(\ref{eq:rot mass}). The radial distance of the gray squares is divided by the $\mu$ from the {\tt baseline} model. The error bar shows 1$\sigma$ uncertainties associated with the range of inclination angles.}
  \label{fig:mass_s0}
\end{figure}

Moreover, we tested the impact of the mask region detected by $^{\tt 3D}${\tt Barolo} on the fitting results. We used different detection modes and noise thresholds to infer the mask region and to reproduce the fitting. The resulting variations in output velocities are minor and within the 1$\sigma$ range above. 

As stated in Sect.~\ref{sec:dyn_mass}, we neglected the effect of pressure support to the modeling, assuming that the gas kinematics of S0 follow perfect circular motions. The low intrinsic velocity dispersion suggests that the ionized gas in S0 is less turbulent than in other disks at $z \gtrsim 1$ \citep[e.g.,][]{swinbank11}, and asymmetric drift corrections due to pressure support should only have minor impact on the measured $V_{\rm rot}$. Other feedback processes could also produce radial motions for the stellar and gas components in S0. In particular, large-scale outflowing winds are a common property for star-forming galaxies at $z \gtrsim 1$ \citep[e.g.,][]{newman12, outflows}, and can influence their internal mass distributions. However, converting the [OII] flux of image S0(d) to a star formation rate, using the \citet{kennicutt98} relation and a \citet{salpeter55} initial mass function, would place S0 near, or even slightly below the star formation main sequence at $z \approx 1.5$ \citep[e.g.,][]{popesso23}. In consequence, while error bars shown in Fig.~\ref{fig:mass_s0} might be broadened by such feedback effects, they are unlikely to play a significant role on the dynamical mass measurement of S0.

\subsection{Uncertainties from strong lens modeling}
\label{subsec:syst_lens}
The uncertainties in the S0 mass measurement through the lensing approach arise from degeneracies between the primary and secondary lens planes, as well as from LoS effects. 

\subsubsection{Two-lens-plane modeling}
\label{subsub:Two lens plane modeling}
The mass distribution of S0 from the two-plane lens models described in Sect.~\ref{sec:lens_mod} can be classified into two groups. Firstly, the model Img$-$2MP based on image positions and the model Img$-$2MP (L/D) combining with the BGG stellar kinematics both predict an elevated mass of $\simeq 10^{11}$\,M$_\odot$ at 3.3~kpc. This falls on the high-mass end of the RC41 sample at $z \simeq 0.65$--2.45 from \citet{Genzel17}. Secondly, the {\tt baseline} model using image positions and the extended arcs as constraints predicts a lower $\theta_{\rm E,S0}$ of 1.65\arcsec. Because the mass degeneracies between the main lens plane at $z = 0.683$, and S0 at $z = 1.49$, are partially broken by the additional constraints from extended arcs, $\theta_{\rm E,S0}$ decreases from $\simeq 4$\arcsec to 1.65\arcsec. 

To further probe this difference, we checked the impact of changing 
the prior on $\theta_{\rm E,S0}$, from the broad flat prior between 0\arcsec\ and 5\arcsec assumed in the {\tt baseline} model, to an informative prior motivated by the kinematic analysis. Assuming that the S0 mass from dynamical and lens modeling are identical, we obtain
\begin{equation}
    \theta_{\text E} = \frac{2\pi}{c^2}V_\text{rot}^2.
    \label{eq:theta_e_Vrot}
\end{equation}
In this case, the best-fit $\theta_{\rm E,S0}$ in the {\tt baseline} model corresponds to a constant $V_{\rm rot}$ of 338.3~km s$^{-1}$ as a function of radius, a value above the 1$\sigma$ upper limit from $^{\tt 3D}${\tt Barolo} (see Fig.~\ref{fig:rotation_curve}). On the other hand, the median rotation velocities correspond to $\theta_{\rm E} \simeq 0.2$\arcsec. We assigned a Gaussian prior $\mathcal{G}~(0.2 \arcsec,\,0.07 \arcsec)$ to $\theta_{\rm E,S0}$ and recomputed the {\tt baseline} model. The best-fit $\theta_{\rm E,S0}$ only decreases mildly to 1.62\arcsec.

Consequently, parameter degeneracies cannot entirely explain the tension between the two modeling approaches. Since the S0 mass is also affected by the choice of the mass parameterization, we also modeled S0 with the elliptical power law and dPIE profiles, finding that the resulting S0 mass remains similar to that of the {\tt baseline}. We thus consider that some additional sources along the LoS discarded from the model might be projected onto the mass plane at $z = 1.49$ and artificially boosting the S0 mass estimate.

\subsubsection{Three-lens-plane modeling}
\label{subsub:Three lens planes modeling}
To further investigate the impact of LoS effects on the S0 mass measurement from strong lensing, we built a model with a third lens plane at $z = 1.36$. Using the MUSE redshift catalog from \citet{Han22}, we find 21 galaxies at $z = 0.7$--0.85 and seven at $z = 1.36$. Two galaxies at $z = 1.36$ are within 15\arcsec\ of the BGG centroid and have F160W $<$ 22\,mag, while none of the galaxies at $z = 0.7$--0.85 reach F160W $<$ 22\,mag within this aperture. Together with the distance to the extended arcs, this suggests that galaxies at $z = 1.36$ are stronger LoS perturbers. We assigned an SIS profile at $z = 1.36$ and performed the lens modeling with three planes at $z = 0.683$ (main lens plane), $z = 1.36$ (perturbers), and $z = 1.49$ (S0), adopting the same constraints as in the {\tt baseline} model. The best-fit model results in $\theta_{\rm E,S0} = 0.85$\arcsec\ and $\theta_{\rm E,z = 1.36} = 0.59$\arcsec\ (Einstein radius of perturbers at $z = 1.36$). 

\begin{figure}
  \centering
\includegraphics[width=1.\linewidth]{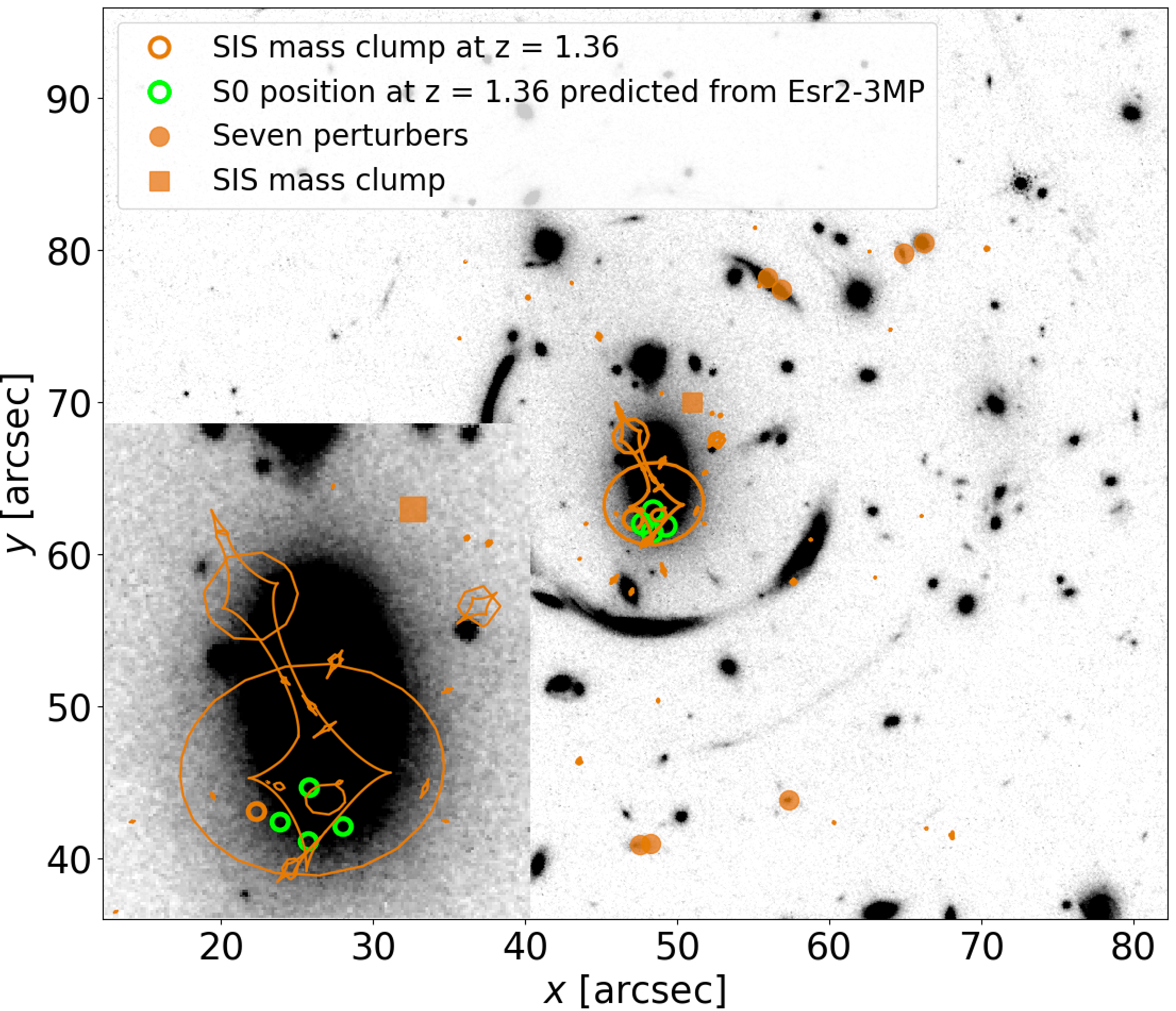}
  \caption{Positions of the seven perturbers at $z = 1.36$ (filled brown circles) and the caustics at $z = 1.36$ based on model Esr2$-$3MP (dashed brown lines). Offsets are relative to the coordinate origin at the bottom-left corner of the HST/F160W image. The filled brown square shows the centroid of the SIS mass profile in model Esr2-3MP. The solid brown and green circles show the centroid of the SIS mass profile and the lensed source S0 projected onto the lens plane at $z = 1.36$, respectively. The position of the SIS at $z = 1.36$ is close that of the projected S0 images, showing that this perturber affects the light propagation of S0 along the LoS.}
  \label{fig:lens_z136_plane}
\end{figure}

The enclosed mass of S0 within 3.3 kpc from this new model (Esr2$-$3MP; see Table~\ref{tab:theta_E_s0}) is close to the 68\% percentile from the kinematic measurement (see the brown line in Fig.~\ref{fig:mass_s0}). We projected the observed image positions of S0 at $z = 0.683$ onto the newly added lens plane at $z = 1.36$, finding that the projected images of S0 are close to the best-fit perturber position (see Fig.~\ref{fig:lens_z136_plane}). The closer the perturber at $z = 1.36$ is to the projected S0 image positions, the greater its best-fit mass. When the perturber moves away from the projected S0 image positions, the value of $\theta_{\rm E,S0}$ falls back to that of the {\tt baseline} model. In the Esr2$-$3MP model, we assigned broad flat priors in the range [0\arcsec, 5\arcsec] for $\theta_{\rm E,S0}$ and $\theta_{\rm E, z=1.36}$, and obtain strong degeneracies between the mass profiles in the three lens planes (see Fig.~\ref{fig:corner_S0_threeplanes}). The $\theta_{\rm E,S0}$ is negatively correlated with the Einstein radius of the group-halo $\theta_{\rm E,GH}$ and with $\theta_{\rm E, z=1.36}$, and is positively correlated with the Einstein radius of the BGG $\theta_{\rm E,BGG}$.
Our model Esr2$-$3MP provides comparable fitting results as {\tt baseline}. It improves the modeling of extended arcs for S0 and S3 by $\mathcal{O}(100)$ in the $\log{\rm likelihood}$\footnote[2]{After marginalizing over the source intensity pixel parameters, i.e., the log likelihood of the lens mass parameters is the log evidence of the source intensity reconstruction given the lens mass parameters}, which indicates the goodness-of-fit \citep{suyu2006} but in the meantime, the $\chi^2$ of image position modeling worsens for S2, S4, and S5 by comparable amounts.

To further explore the mass distribution at $z = 1.36$, we performed the three-lens-plane modeling with either a constant mass sheet (Esr2$-$3MP (MS)) or a constant mass sheet with an external shear (Esr2$-$3MP (MS + Shear)) replacing the SIS. The former shows similar fitting results as Esr2$-$3MP, while the latter improves both the image position and extended image modeling by $\mathcal{O}(100)$. Notably, Esr2$-$3MP (MS + Shear) significantly improves the image position fitting of the multiple-image set S4 at $z = 4.21$. However, this model has nearly negligible mass on the mass sheet at $z = 1.36$, resulting in a substantial S0 mass prediction with $\theta_{E,S0} = 2.01 \arcsec$ (see Table~\ref{tab:theta_E_s02}); this translates to $V_{\rm rot} = 382$ km s$^{-1}$, which is higher than the $V_{\rm rot}$ inferred from the kinematic model. The critical curves for the lens plane at $z = 1.36$ in these three models almost completely overlap while the caustics exhibit similar shapes but are scaled differently (see Figs.~\ref{fig:crit3} and \ref{fig:caus3}). The rescaling of caustics between Esr2$-$3MP and Esr2$-$3MP (MS) can be explained by a generalized mass-sheet transformation in a multi-plane lens scenario \citep{Schneidermulti} that leaves all observables but the time delay invariant, corresponding to a uniform isotropic scaling in each lens and source plane. It is also not surprising that the critical and caustic curves of Esr2$-$3MP (MS) and Esr2$-$3MP (MS + Shear) for lens plane at $z = 1.36$ are nearly identical because the Esr2$-$3MP (MS) yields a best-fit mass sheet with 0.03 and Esr2$-$3MP (MS + Shear) has essentially zero mass sheet and a minor external shear with the strength of 0.04. In the Esr2$-$3MP (MS + Shear) model, the best-fit position angles of the external shear at $ z = 1.36$ and the main lens plane at $z =0.683$ are $49.0^\circ$ and $-37.9^\circ$ counterclockwise from the x-axis, respectively. These directions of mass concentrations from the shear are consistent with the overdensity of galaxies in the environment of CSWA\,31 (see Fig.~\ref{fig:crit3}).

These models can be considered as additional tests illustrating the differences in the predicted S0 mass when adding a third lens plane. Overall, the model Esr2$-$3MP provides comparable fitting results as {\tt baseline} and is also consistent with the kinematics modeling. In this sense, a concentrated mass profile performs better than adding a global change in the third plane. Nevertheless, the value of $\theta_{\rm E, z=1.36} = 0.59\arcsec$ is lower than for typical dark-matter halos and the best-fit perturber centroid does not fall in the vicinity of HST-detected galaxies. This shows that the actual LoS structure in CSWA\,31 is complex and cannot be fully captured by simple mass profiles, i,e., SIS, constant mass sheet or shear. Possible asymmetries in the overall group-scale mass distribution could also play a role.

\section{Summary}
\label{sec:summary}
In this work we have explored the constraints on the total masses of secondary lenses obtained from multi-plane strong lensing modeling. We performed a case study of a lensed spiral galaxy at $z = 1.49$ in CSWA\,31, which is along the LoS to the BGG and its group halo at $z = 0.683$. This system forms a rare, well-suited configuration with multiple image sets at different redshifts, providing tight constraints on the lens model and adequate data to infer the maximum circular velocity of the spiral.

We measured the total mass of S0 from independent strong lensing and kinematic modeling in order to assess the robustness of secondary lens mass measurements (see Sect.~\ref{sec:discu}). Relying on the detailed strong lensing analysis of \citet{Han22}, we measured the S0 mass with a range of models based on different constraints. We also modeled the velocity gradient using the least distorted image S0(d) with $\mu \simeq 7$, using the state-of-the-art software $^{\tt 3D}${\tt Barolo}. We obtain a median $V_{\rm rot} = 123_{-38.3}^{+117}$ng at 3.3 kpc, consistent with other kinematic models of high-redshift disks of similar stellar masses \citep[e.g.,][]{Genzel17,Genzel2020}. The total masses of S0 derived via the two approaches are only compatible when adding a third lens plane with a concentrated mass distribution to account for LoS perturbers.

We conclude that the secondary-lens measurement from the multiple-plane modeling includes high systematic uncertainties from degeneracies between lens planes and the complex LoS structure. Leveraging extended arcs as additional constraints can reduce the degeneracies. The secondary lens contributes less to the deflection angle of light from the source galaxies compared to the main lens, and its best-fit properties can easily be affected by other perturbers along the LoS since they project their mass onto the secondary lens plane. Accounting for additional structures along the LoS is therefore key to improving the mass measurement of the secondary lens.

\begin{acknowledgements}

We thank E.~Di Teodoro for helpful discussions and support on $^{\tt 3D}${\tt Barolo}, and the anonymous referee for the constructive feedback that improved our paper. HW, RC and SHS thank the Max Planck Society for support through the Max Planck Research Group and Max Planck Fellowship for SHS. AG acknowledges funding and support by the Swiss National Science Foundation (SNSF). C.G. acknowledges support through grant MIUR2020 SKSTHZ. 

\end{acknowledgements}

\bibliographystyle{aa}
\bibliography{reference}

\begin{thebibliography}{45}
\expandafter\ifx\csname natexlab\endcsname\relax\def\natexlab#1{#1}\fi

\bibitem[{{Auger} {et~al.}(2010){Auger}, {Treu}, {Bolton}, {Gavazzi}, {Koopmans}, {Marshall}, {Moustakas}, \& {Burles}}]{auger10}
{Auger}, M.~W., {Treu}, T., {Bolton}, A.~S., {et~al.} 2010, \apj, 724, 511

\bibitem[{{Bacon} {et~al.}(2015){Bacon}, {Brinchmann}, {Richard}, {Contini}, {Drake}, {Franx}, {Tacchella}, {Vernet}, {Wisotzki}, {Blaizot}, {Bouch{\'e}}, {Bouwens}, {Cantalupo}, {Carollo}, {Carton}, {Caruana}, {Cl{\'e}ment}, {Dreizler}, {Epinat}, {Guiderdoni}, {Herenz}, {Husser}, {Kamann}, {Kerutt}, {Kollatschny}, {Krajnovic}, {Lilly}, {Martinsson}, {Michel-Dansac}, {Patricio}, {Schaye}, {Shirazi}, {Soto}, {Soucail}, {Steinmetz}, {Urrutia}, {Weilbacher}, \& {de Zeeuw}}]{muse}
{Bacon}, R., {Brinchmann}, J., {Richard}, J., {et~al.} 2015, \aap, 575, A75

\bibitem[{{Barkana}(1998)}]{spemd}
{Barkana}, R. 1998, \apj, 502, 531

\bibitem[{{Bayliss} {et~al.}(2014){Bayliss}, {Johnson}, {Gladders}, {Sharon}, \& {Oguri}}]{Bayliss14}
{Bayliss}, M.~B., {Johnson}, T., {Gladders}, M.~D., {Sharon}, K., \& {Oguri}, M. 2014, \apj, 783, 41

\bibitem[{{Belokurov} {et~al.}(2009){Belokurov}, {Evans}, {Hewett}, {Moiseev}, {McMahon}, {Sanchez}, \& {King}}]{belokurov09}
{Belokurov}, V., {Evans}, N.~W., {Hewett}, P.~C., {et~al.} 2009, \mnras, 392, 104

\bibitem[{{Birrer} {et~al.}(2019){Birrer}, {Treu}, {Rusu}, {Bonvin}, {Fassnacht}, {Chan}, {Agnello}, {Shajib}, {Chen}, {Auger}, {Courbin}, {Hilbert}, {Sluse}, {Suyu}, {Wong}, {Marshall}, {Lemaux}, \& {Meylan}}]{Birrer19}
{Birrer}, S., {Treu}, T., {Rusu}, C.~E., {et~al.} 2019, \mnras, 484, 4726

\bibitem[{{Boquien} {et~al.}(2019){Boquien}, {Burgarella}, {Roehlly}, {Buat}, {Ciesla}, {Corre}, {Inoue}, \& {Salas}}]{boquien19}
{Boquien}, M., {Burgarella}, D., {Roehlly}, Y., {et~al.} 2019, \aap, 622, A103

\bibitem[{{Bournaud} {et~al.}(2009){Bournaud}, {Elmegreen}, \& {Martig}}]{turbulentdisk2009}
{Bournaud}, F., {Elmegreen}, B.~G., \& {Martig}, M. 2009, \apjl, 707, L1

\bibitem[{{Bradshaw} {et~al.}(2013){Bradshaw}, {Almaini}, {Hartley}, {Smith}, {Conselice}, {Dunlop}, {Simpson}, {Chuter}, {Cirasuolo}, {Foucaud}, {McLure}, {Mortlock}, \& {Pearce}}]{outflows}
{Bradshaw}, E.~J., {Almaini}, O., {Hartley}, W.~G., {et~al.} 2013, \mnras, 433, 194

\bibitem[{{Chiriv{\`\i}} {et~al.}(2018){Chiriv{\`\i}}, {Suyu}, {Grillo}, {Halkola}, {Balestra}, {Caminha}, {Mercurio}, \& {Rosati}}]{Chirivi2018}
{Chiriv{\`\i}}, G., {Suyu}, S.~H., {Grillo}, C., {et~al.} 2018, \aap, 614, A8

\bibitem[{{Collett}(2015)}]{Collett+2015}
{Collett}, T.~E. 2015, \apj, 811, 20

\bibitem[{{Di Teodoro} \& {Fraternali}(2015)}]{3Dbarolo}
{Di Teodoro}, E.~M. \& {Fraternali}, F. 2015, \mnras, 451, 3021

\bibitem[{{Di Teodoro} {et~al.}(2018){Di Teodoro}, {Grillo}, {Fraternali}, {Gobat}, {Karman}, {Mercurio}, {Rosati}, {Balestra}, {Caminha}, {Caputi}, {Lombardi}, {Suyu}, {Treu}, \& {Vanzella}}]{diteodoro18}
{Di Teodoro}, E.~M., {Grillo}, C., {Fraternali}, F., {et~al.} 2018, \mnras, 476, 804

\bibitem[{{El{\'\i}asd{\'o}ttir} {et~al.}(2007){El{\'\i}asd{\'o}ttir}, {Limousin}, {Richard}, {Hjorth}, {Kneib}, {Natarajan}, {Pedersen}, {Jullo}, \& {Paraficz}}]{dpie2007}
{El{\'\i}asd{\'o}ttir}, {\'A}., {Limousin}, M., {Richard}, J., {et~al.} 2007, arXiv e-prints, arXiv:0710.5636

\bibitem[{{Gavazzi} {et~al.}(2008){Gavazzi}, {Treu}, {Koopmans}, {Bolton}, {Moustakas}, {Burles}, \& {Marshall}}]{doubellens}
{Gavazzi}, R., {Treu}, T., {Koopmans}, L. V.~E., {et~al.} 2008, \apj, 677, 1046

\bibitem[{{Genzel} {et~al.}(2017){Genzel}, {F{\"o}rster Schreiber}, {{\"U}bler}, {Lang}, {Naab}, {Bender}, {Tacconi}, {Wisnioski}, {Wuyts}, {Alexander}, {Beifiori}, {Belli}, {Brammer}, {Burkert}, {Carollo}, {Chan}, {Davies}, {Fossati}, {Galametz}, {Genel}, {Gerhard}, {Lutz}, {Mendel}, {Momcheva}, {Nelson}, {Renzini}, {Saglia}, {Sternberg}, {Tacchella}, {Tadaki}, \& {Wilman}}]{Genzel17}
{Genzel}, R., {F{\"o}rster Schreiber}, N.~M., {{\"U}bler}, H., {et~al.} 2017, \nat, 543, 397

\bibitem[{{Genzel} {et~al.}(2020){Genzel}, {Price}, {{\"U}bler}, {F{\"o}rster Schreiber}, {Shimizu}, {Tacconi}, {Bender}, {Burkert}, {Contursi}, {Coogan}, {Davies}, {Davies}, {Dekel}, {Herrera-Camus}, {Lee}, {Lutz}, {Naab}, {Neri}, {Nestor}, {Renzini}, {Saglia}, {Schuster}, {Sternberg}, {Wisnioski}, \& {Wuyts}}]{Genzel2020}
{Genzel}, R., {Price}, S.~H., {{\"U}bler}, H., {et~al.} 2020, \apj, 902, 98

\bibitem[{{Grillo} {et~al.}(2015){Grillo}, {Suyu}, {Rosati}, {Mercurio}, {Balestra}, {Munari}, {Nonino}, {Caminha}, {Lombardi}, {De Lucia}, {Borgani}, {Gobat}, {Biviano}, {Girardi}, {Umetsu}, {Coe}, {Koekemoer}, {Postman}, {Zitrin}, {Halkola}, {Broadhurst}, {Sartoris}, {Presotto}, {Annunziatella}, {Maier}, {Fritz}, {Vanzella}, \& {Frye}}]{Grillo2015}
{Grillo}, C., {Suyu}, S.~H., {Rosati}, P., {et~al.} 2015, \apj, 800, 38

\bibitem[{{Kennicutt}(1998)}]{kennicutt98}
{Kennicutt}, Robert~C., J. 1998, \araa, 36, 189

\bibitem[{{Koopmans} {et~al.}(2006){Koopmans}, {Treu}, {Bolton}, {Burles}, \& {Moustakas}}]{koopmans06}
{Koopmans}, L. V.~E., {Treu}, T., {Bolton}, A.~S., {Burles}, S., \& {Moustakas}, L.~A. 2006, \apj, 649, 599

\bibitem[{Mandelbaum {et~al.}(2018)Mandelbaum, Eifler, Hložek, Collett, Gawiser, Scolnic, Alonso, Awan, Biswas, Blazek, Burchat, Chisari, Dell'Antonio, Digel, Frieman, Goldstein, Hook, Željko Ivezić, Kahn, Kamath, Kirkby, Kitching, Krause, Leget, Marshall, Meyers, Miyatake, Newman, Nichol, Rykoff, Sanchez, Slosar, Sullivan, \& Troxel}]{mandelbaum18}
Mandelbaum, R., Eifler, T., Hložek, R., {et~al.} 2018, The LSST Dark Energy Science Collaboration (DESC) Science Requirements Document

\bibitem[{{McGaugh} \& {Schombert}(2015)}]{Tullyfisher}
{McGaugh}, S.~S. \& {Schombert}, J.~M. 2015, \apj, 802, 18

\bibitem[{{Meneghetti} {et~al.}(2017){Meneghetti}, {Natarajan}, {Coe}, {Contini}, {De Lucia}, {Giocoli}, {Acebron}, {Borgani}, {Bradac}, {Diego}, {Hoag}, {Ishigaki}, {Johnson}, {Jullo}, {Kawamata}, {Lam}, {Limousin}, {Liesenborgs}, {Oguri}, {Sebesta}, {Sharon}, {Williams}, \& {Zitrin}}]{Meneghetti}
{Meneghetti}, M., {Natarajan}, P., {Coe}, D., {et~al.} 2017, \mnras, 472, 3177

\bibitem[{{Nestor Shachar} {et~al.}(2023){Nestor Shachar}, {Price}, {F{\"o}rster Schreiber}, {Genzel}, {Shimizu}, {Tacconi}, {{\"U}bler}, {Burkert}, {Davies}, {Dekel}, {Herrera-Camus}, {Lee}, {Liu}, {Lutz}, {Naab}, {Neri}, {Renzini}, {Saglia}, {Schuster}, {Sternberg}, {Wisnioski}, \& {Wuyts}}]{RC100}
{Nestor Shachar}, A., {Price}, S.~H., {F{\"o}rster Schreiber}, N.~M., {et~al.} 2023, \apj, 944, 78

\bibitem[{{Newman} {et~al.}(2012){Newman}, {Genzel}, {F{\"o}rster-Schreiber}, {Shapiro Griffin}, {Mancini}, {Lilly}, {Renzini}, {Bouch{\'e}}, {Burkert}, {Buschkamp}, {Carollo}, {Cresci}, {Davies}, {Eisenhauer}, {Genel}, {Hicks}, {Kurk}, {Lutz}, {Naab}, {Peng}, {Sternberg}, {Tacconi}, {Vergani}, {Wuyts}, \& {Zamorani}}]{newman12}
{Newman}, S.~F., {Genzel}, R., {F{\"o}rster-Schreiber}, N.~M., {et~al.} 2012, \apj, 761, 43

\bibitem[{{Peng} {et~al.}(2010){Peng}, {Lilly}, {Kova{\v{c}}}, {Bolzonella}, {Pozzetti}, {Renzini}, {Zamorani}, {Ilbert}, {Knobel}, {Iovino}, {Maier}, {Cucciati}, {Tasca}, {Carollo}, {Silverman}, {Kampczyk}, {de Ravel}, {Sanders}, {Scoville}, {Contini}, {Mainieri}, {Scodeggio}, {Kneib}, {Le F{\`e}vre}, {Bardelli}, {Bongiorno}, {Caputi}, {Coppa}, {de la Torre}, {Franzetti}, {Garilli}, {Lamareille}, {Le Borgne}, {Le Brun}, {Mignoli}, {Perez Montero}, {Pello}, {Ricciardelli}, {Tanaka}, {Tresse}, {Vergani}, {Welikala}, {Zucca}, {Oesch}, {Abbas}, {Barnes}, {Bordoloi}, {Bottini}, {Cappi}, {Cassata}, {Cimatti}, {Fumana}, {Hasinger}, {Koekemoer}, {Leauthaud}, {Maccagni}, {Marinoni}, {McCracken}, {Memeo}, {Meneux}, {Nair}, {Porciani}, {Presotto}, \& {Scaramella}}]{2010ApJ...721..193P}
{Peng}, Y.-j., {Lilly}, S.~J., {Kova{\v{c}}}, K., {et~al.} 2010, \apj, 721, 193

\bibitem[{{Popesso} {et~al.}(2023){Popesso}, {Concas}, {Cresci}, {Belli}, {Rodighiero}, {Inami}, {Dickinson}, {Ilbert}, {Pannella}, \& {Elbaz}}]{popesso23}
{Popesso}, P., {Concas}, A., {Cresci}, G., {et~al.} 2023, \mnras, 519, 1526

\bibitem[{{Price} {et~al.}(2021){Price}, {Shimizu}, {Genzel}, {{\"U}bler}, {F{\"o}rster Schreiber}, {Tacconi}, {Davies}, {Coogan}, {Lutz}, {Wuyts}, {Wisnioski}, {Nestor}, {Sternberg}, {Burkert}, {Bender}, {Contursi}, {Davies}, {Herrera-Camus}, {Lee}, {Naab}, {Neri}, {Renzini}, {Saglia}, {Schruba}, \& {Schuster}}]{price21}
{Price}, S.~H., {Shimizu}, T.~T., {Genzel}, R., {et~al.} 2021, \apj, 922, 143

\bibitem[{{Puglisi} {et~al.}(2023){Puglisi}, {Dudzevi{\v{c}}i{\={u}}t{\.{e}}}, {Swinbank}, {Gillman}, {Tiley}, {Bower}, {Cirasuolo}, {Cortese}, {Glazebrook}, {Harrison}, {Ibar}, {Molina}, {Obreschkow}, {Oman}, {Schaller}, {Shankar}, \& {Sharples}}]{Puglisispiral}
{Puglisi}, A., {Dudzevi{\v{c}}i{\={u}}t{\.{e}}}, U., {Swinbank}, M., {et~al.} 2023, \mnras, 524, 2814

\bibitem[{{Salpeter}(1955)}]{salpeter55}
{Salpeter}, E.~E. 1955, \apj, 121, 161

\bibitem[{{Schneider}(2014)}]{Schneidermulti}
{Schneider}, P. 2014, arXiv e-prints, arXiv:1409.0015

\bibitem[{{Shajib} {et~al.}(2022){Shajib}, {Vernardos}, {Collett}, {Motta}, {Sluse}, {Williams}, {Saha}, {Birrer}, {Spiniello}, \& {Treu}}]{Anowar}
{Shajib}, A.~J., {Vernardos}, G., {Collett}, T.~E., {et~al.} 2022, arXiv e-prints, arXiv:2210.10790

\bibitem[{{Soto} {et~al.}(2012){Soto}, {Martin}, {Prescott}, \& {Armus}}]{shockoutflow}
{Soto}, K.~T., {Martin}, C.~L., {Prescott}, M.~K.~M., \& {Armus}, L. 2012, \apj, 757, 86

\bibitem[{{Stark} {et~al.}(2013){Stark}, {Auger}, {Belokurov}, {Jones}, {Robertson}, {Ellis}, {Sand}, {Moiseev}, {Eagle}, \& {Myers}}]{stark13}
{Stark}, D.~P., {Auger}, M., {Belokurov}, V., {et~al.} 2013, \mnras, 436, 1040

\bibitem[{{Suyu} \& {Halkola}(2010{\natexlab{a}})}]{suyu2010}
{Suyu}, S.~H. \& {Halkola}, A. 2010{\natexlab{a}}, \aap, 524, A94

\bibitem[{{Suyu} \& {Halkola}(2010{\natexlab{b}})}]{dpie2010}
{Suyu}, S.~H. \& {Halkola}, A. 2010{\natexlab{b}}, \aap, 524, A94

\bibitem[{{Suyu} {et~al.}(2012){Suyu}, {Hensel}, {McKean}, {Fassnacht}, {Treu}, {Halkola}, {Norbury}, {Jackson}, {Schneider}, {Thompson}, {Auger}, {Koopmans}, \& {Matthews}}]{Suyu2012}
{Suyu}, S.~H., {Hensel}, S.~W., {McKean}, J.~P., {et~al.} 2012, \apj, 750, 10

\bibitem[{{Suyu} {et~al.}(2006){Suyu}, {Marshall}, {Hobson}, \& {Blandford}}]{suyu2006}
{Suyu}, S.~H., {Marshall}, P.~J., {Hobson}, M.~P., \& {Blandford}, R.~D. 2006, \mnras, 371, 983

\bibitem[{{Swinbank} {et~al.}(2011){Swinbank}, {Papadopoulos}, {Cox}, {Krips}, {Ivison}, {Smail}, {Thomson}, {Neri}, {Richard}, \& {Ebeling}}]{swinbank11}
{Swinbank}, A.~M., {Papadopoulos}, P.~P., {Cox}, P., {et~al.} 2011, \apj, 742, 11

\bibitem[{{Treu}(2010)}]{Tommaso}
{Treu}, T. 2010, \araa, 48, 87

\bibitem[{{{\"U}bler} {et~al.}(2019){{\"U}bler}, {Genzel}, {Wisnioski}, {F{\"o}rster Schreiber}, {Shimizu}, {Price}, {Tacconi}, {Belli}, {Wilman}, {Fossati}, {Mendel}, {Davies}, {Beifiori}, {Bender}, {Brammer}, {Burkert}, {Chan}, {Davies}, {Fabricius}, {Galametz}, {Herrera-Camus}, {Lang}, {Lutz}, {Momcheva}, {Naab}, {Nelson}, {Saglia}, {Tadaki}, {van Dokkum}, \& {Wuyts}}]{Ubler19}
{{\"U}bler}, H., {Genzel}, R., {Wisnioski}, E., {et~al.} 2019, \apj, 880, 48

\bibitem[{{Wang} {et~al.}(2022){Wang}, {Ca{\~n}ameras}, {Caminha}, {Suyu}, {Y{\i}ld{\i}r{\i}m}, {Chiriv{\`\i}}, {Christensen}, {Grillo}, \& {Schuldt}}]{Han22}
{Wang}, H., {Ca{\~n}ameras}, R., {Caminha}, G.~B., {et~al.} 2022, \aap, 668, A162

\bibitem[{{Wisnioski} {et~al.}(2015){Wisnioski}, {F{\"o}rster Schreiber}, {Wuyts}, {Wuyts}, {Bandara}, {Wilman}, {Genzel}, {Bender}, {Davies}, {Fossati}, {Lang}, {Mendel}, {Beifiori}, {Brammer}, {Chan}, {Fabricius}, {Fudamoto}, {Kulkarni}, {Kurk}, {Lutz}, {Nelson}, {Momcheva}, {Rosario}, {Saglia}, {Seitz}, {Tacconi}, \& {van Dokkum}}]{Wisnioski15}
{Wisnioski}, E., {F{\"o}rster Schreiber}, N.~M., {Wuyts}, S., {et~al.} 2015, \apj, 799, 209

\bibitem[{{Wong} {et~al.}(2017){Wong}, {Suyu}, {Auger}, {Bonvin}, {Courbin}, {Fassnacht}, {Halkola}, {Rusu}, {Sluse}, {Sonnenfeld}, {Treu}, {Collett}, {Hilbert}, {Koopmans}, {Marshall}, \& {Rumbaugh}}]{Wong+17}
{Wong}, K.~C., {Suyu}, S.~H., {Auger}, M.~W., {et~al.} 2017, \mnras, 465, 4895

\bibitem[{{Yu} {et~al.}(2022){Yu}, {Xu}, {Ho}, {Wang}, \& {Kao}}]{secularinst}
{Yu}, S.-Y., {Xu}, D., {Ho}, L.~C., {Wang}, J., \& {Kao}, W.-B. 2022, \aap, 661, A98

\end{thebibliography}

\begin{appendix}

\section{Kinematics of S0}
\begin{figure}
  \centering  \includegraphics[width=1.\linewidth]{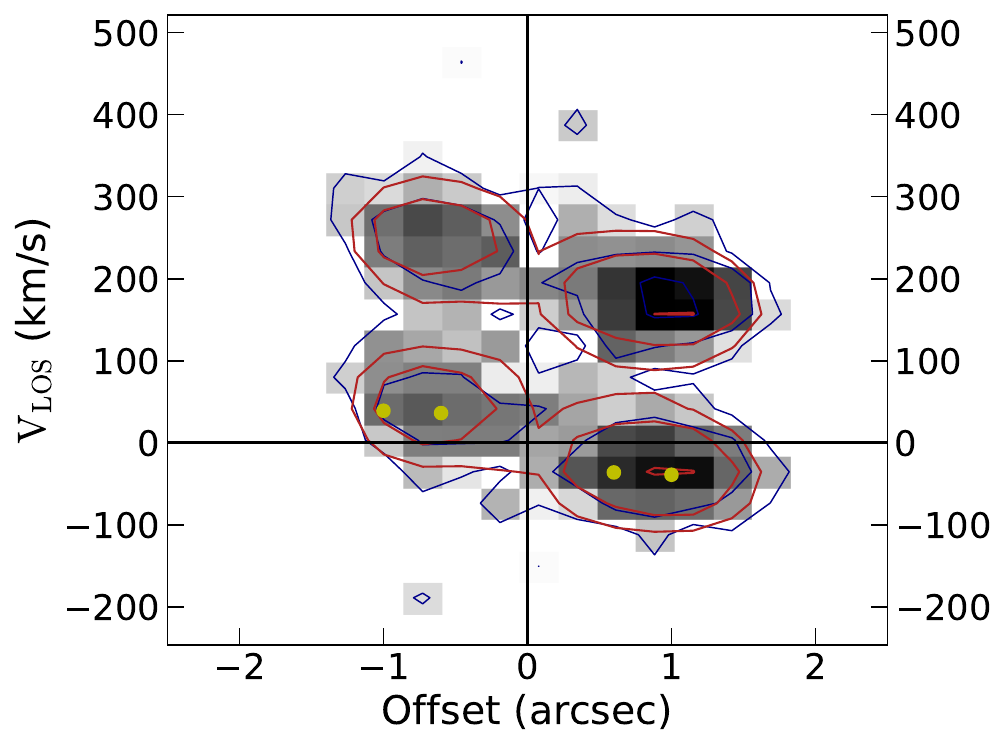}
  \caption{Position-velocity diagram along the major axis of S0(d) for the [OII] doublet. The contour levels and color coding are the same as in Fig.~\ref{fig:channel_map.pdf}. The yellow points show the position on the S0(d) image where we have secure velocity measurements. We 
  set the $V_{\rm LoS} = 0$~km s$^{-1}$ at the center of [OII]$\lambda$3726~$\AA$ emission line. }
  \label{fig:pv_major}
\end{figure}

\begin{figure}
  \centering
  \includegraphics[width=1.\linewidth]{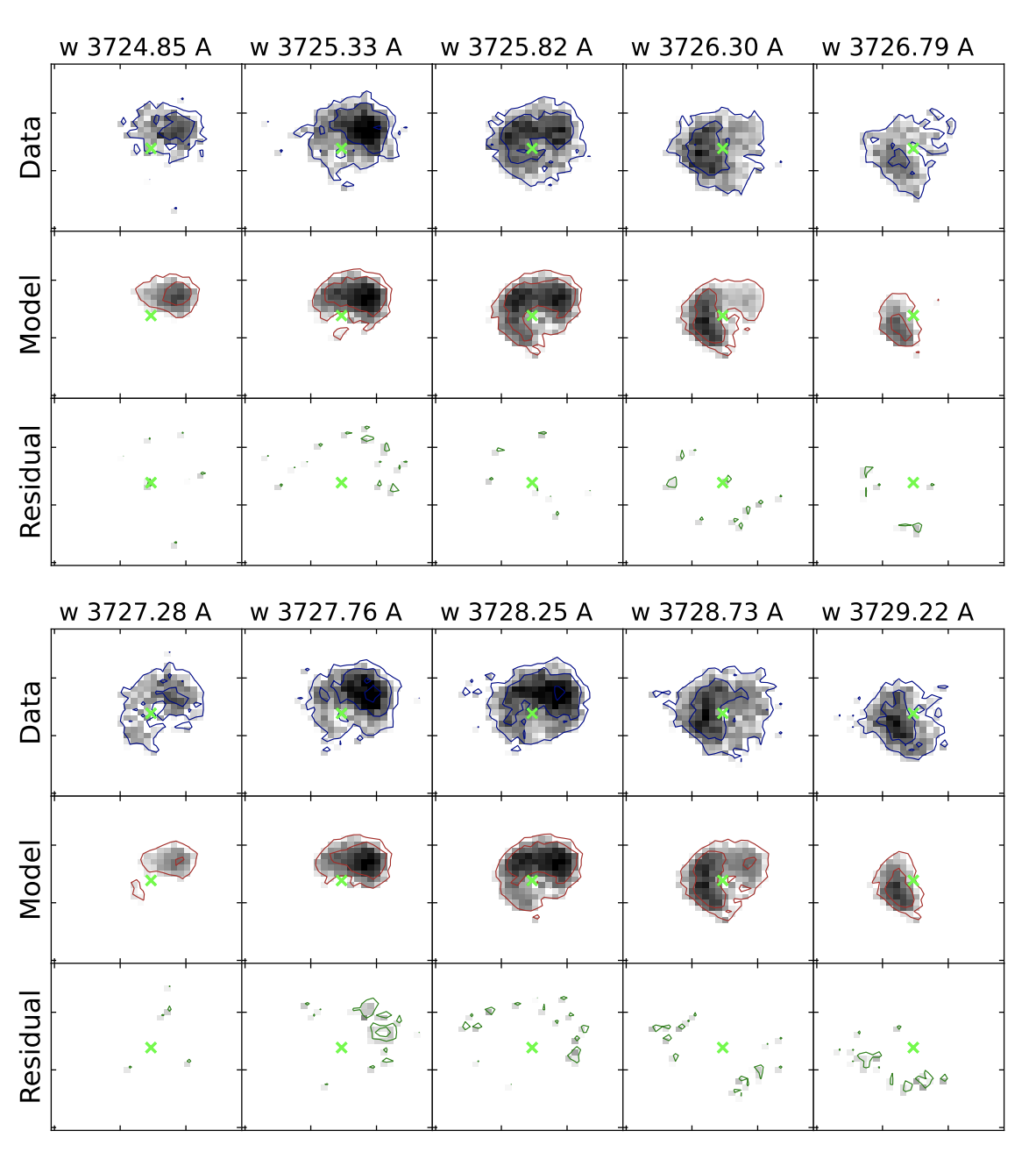}
  \caption{Channel maps for the kinematic models with three rings. The contour colors and levels are the same as in Fig.~\ref{fig:channel_map.pdf}. The residual of the [OII]$\lambda$3729~$\AA$ emission line (bottom row) is worse than the models with four rings. }
  \label{fig:channel_map_three_rings}
\end{figure}

\begin{figure}
  \centering  \includegraphics[width=1.\linewidth]{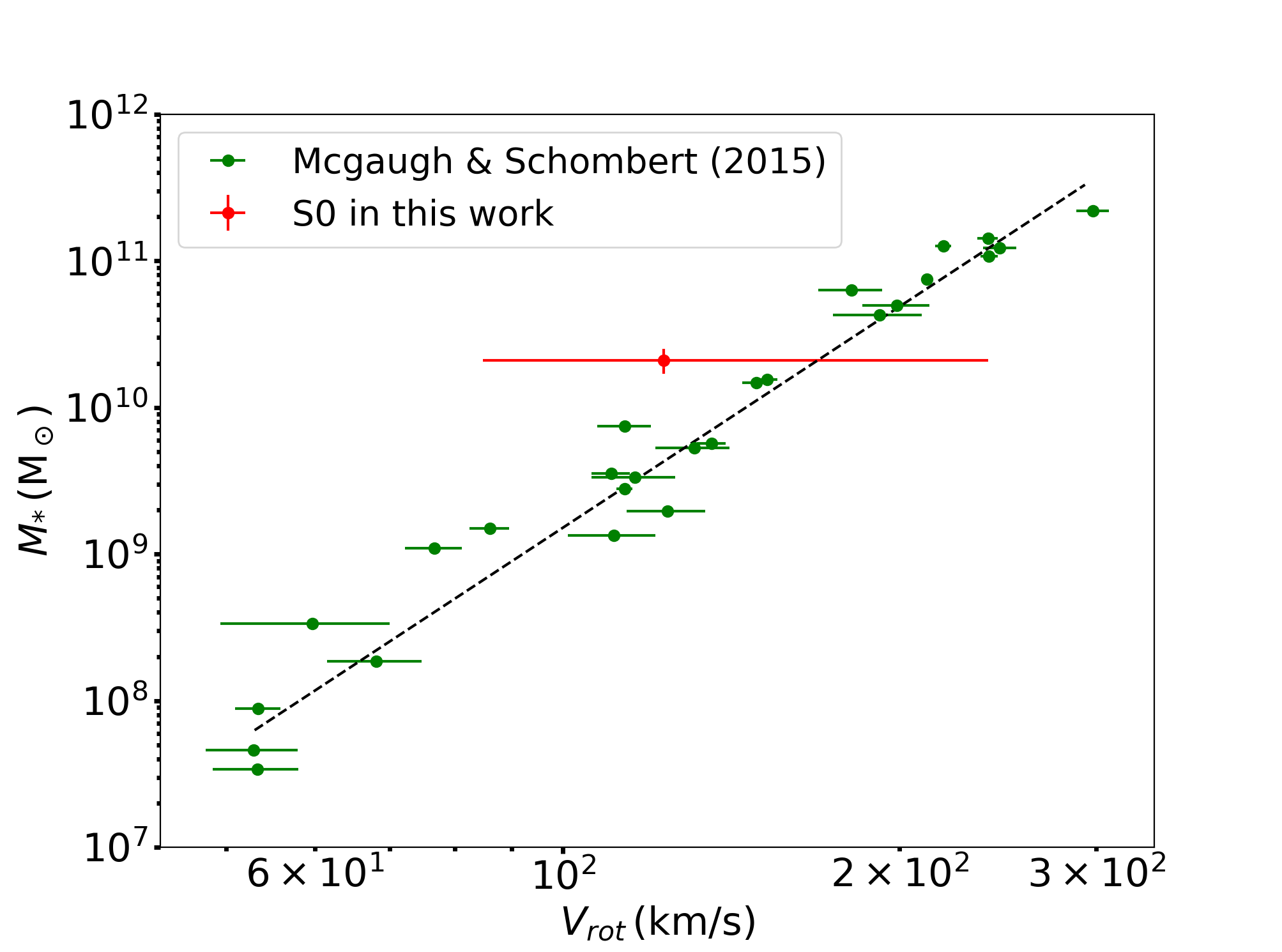}
  \caption{Stellar-mass TFR. The green points show the sample of galaxies at $z \simeq 0$ from \citet[their Table~7]{Tullyfisher}, and the dashed line shows the corresponding best-fitting relation. The red point shows the S0 spiral at $z=1.49$ from this work with 1$\sigma$ uncertainty, which is consistent with the local TFR.}
  \label{fig:TFR}
\end{figure}

We present the kinematic modeling results from $^{\tt 3D}${\tt Barolo} and the stellar-mass TFR in Figs.~\ref{fig:pv_major}, \ref{fig:channel_map_three_rings}, and \ref{fig:TFR}.

\section{Three-lens-plane models}
\begin{figure}
  \centering
  \includegraphics[width=1.\linewidth]{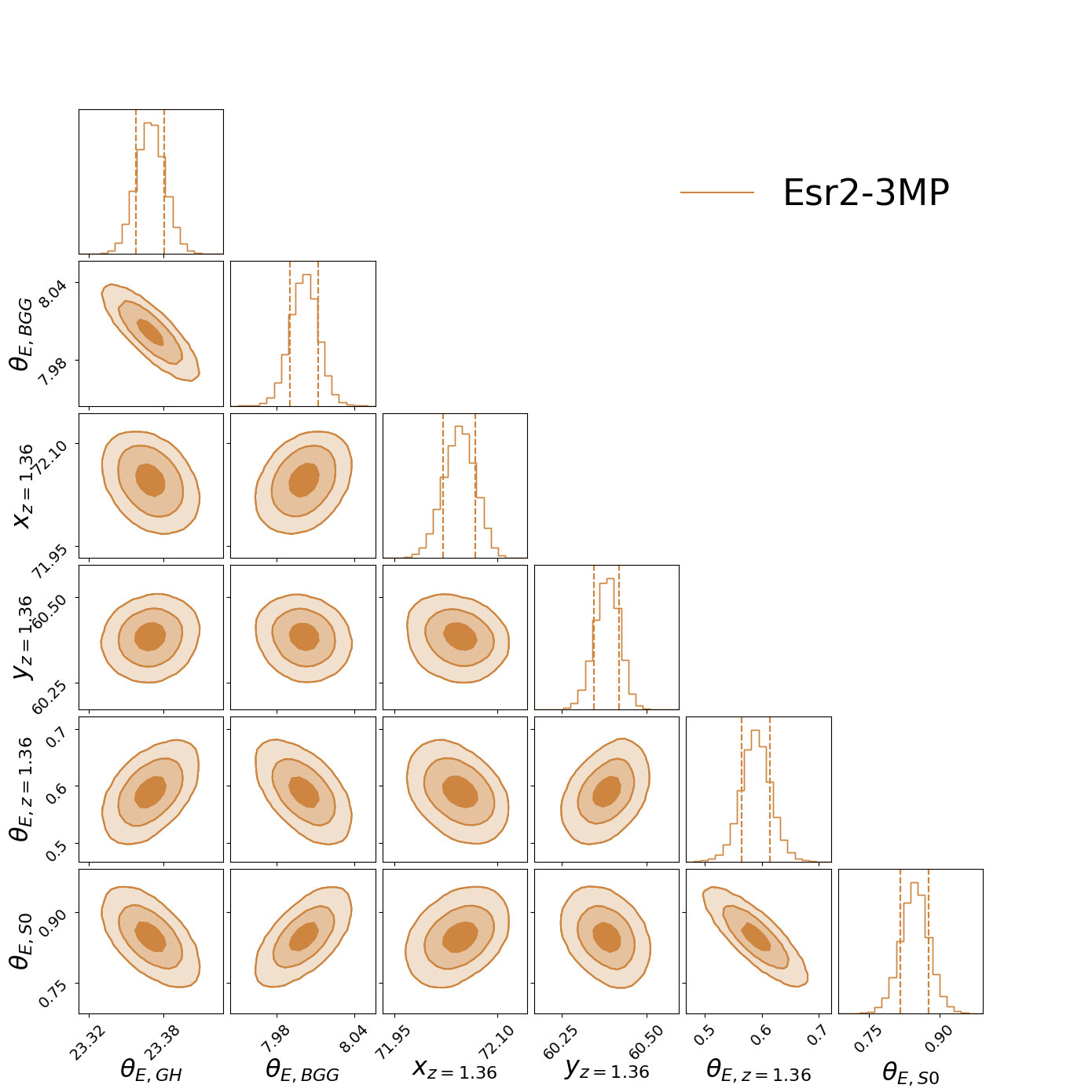}
  \caption{Joint posterior PDFs for the model Esr2-3MP based on two extended sources and image positions. We show the parameters representing the lens masses at three multiple planes. The $\theta_{\rm E,GH}$ and $\theta_{\rm E,BGG}$ are the Einstein radius of the group halo and the BGG at the main lens plane at $z=0.683$. The $x_{\rm z=1.36}$ and $y_{\rm z=1.36}$ are the centroid of the perturber, and the $\theta_{\rm E, z=1.36}$ is its Einstein radius. The $\theta_{\rm E, S0}$ is the Einstein radius of S0 at $z=1.49$. The three shaded areas on the joint PDFs show the $68.3\%$, $95.4\%$, and $99.7\%$ credible regions. The 1D histograms show the marginalized PDFs for the selected mass parameters, and the vertical lines mark the 1$\sigma$ confidence intervals. The $\theta_{\rm E, S0}$ shows strong degeneracies with $\theta_{\rm E,BGG}$, $\theta_{\rm E,GH}$, and $\theta_{\rm E, z=1.36}$. }
  \label{fig:corner_S0_threeplanes}
\end{figure}

\begin{figure}
  \centering
  \includegraphics[width=1.1\linewidth]{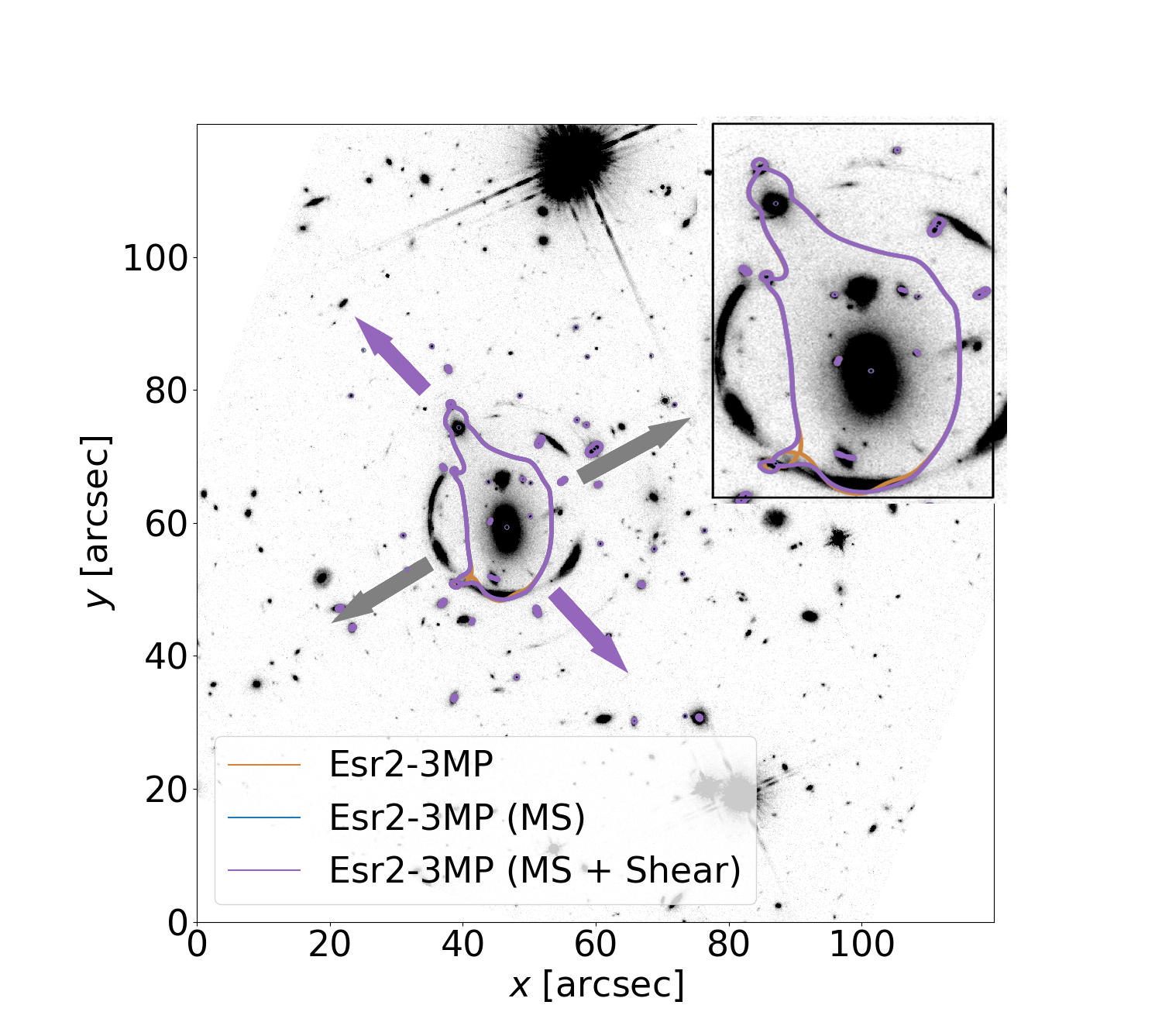}
  \caption{Critical curves for the lens plane at $z = 1.36$ in three-lens-plane models. The Esr2$-$3MP model uses an isothermal mass profile to model the mass distribution at $z = 1.36$. Meanwhile, the Esr2$-$3MP (MS) model uses a constant mass sheet, and the Esr2$-$3MP (MS + Shear) model uses a constant mass sheet with an external shear. The purple and gray arrows indicate the mass concentrations predicted by the external shear at $z = 1.36$ and $z = 0.683$, respectively, in the model Esr2$-$3MP (MS + Shear). The critical curves for these models nearly overlap, with the orange and blue lines largely hidden beneath the purple line.}
  \label{fig:crit3}
\end{figure}

\begin{figure}
  \centering
  \includegraphics[width=0.8\linewidth]{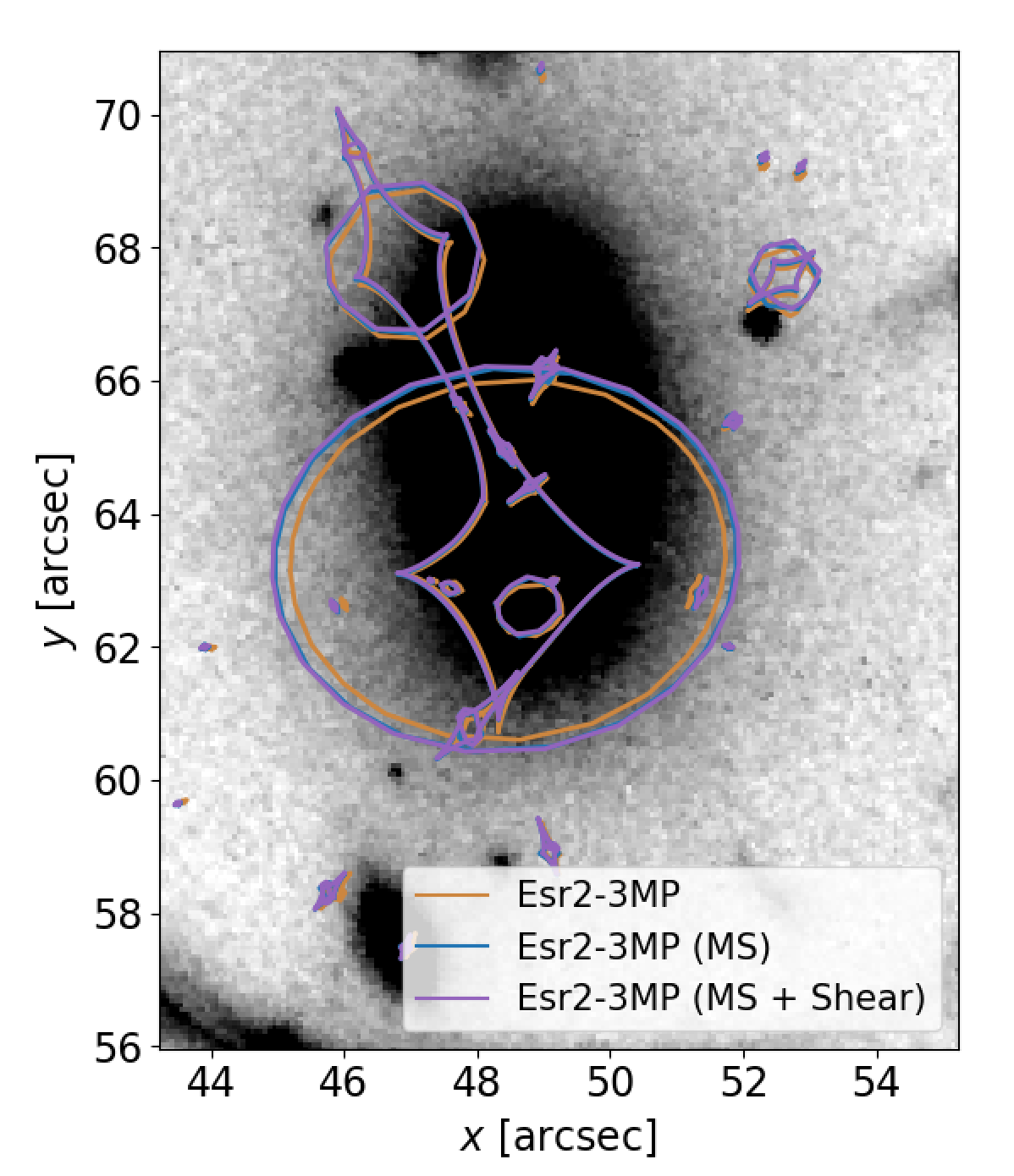}
  \caption{Caustics for lens plane at $z = 1.36$ in three-lens-plane models.}
  \label{fig:caus3}
\end{figure}

We present the degeneracies of the mass parameters in model Esr-3MP in Fig.~\ref{fig:corner_S0_threeplanes}. We list the best-fit parameters in two complementary models Esr2$-$3MP (MS) and Esr$-$3MP (MS + Shear) in Table~\ref{tab:theta_E_s02}
and we show their critical curves and caustics for the lens plane at $z = 1.36$ in Figs.~\ref{fig:crit3} and ~\ref{fig:caus3}.

\begin{table}
  \caption{Complementary three-lens-plane models with the constant mass sheet at $z = 1.36$, their constraints, the best-fit values of $\theta_{\rm E,S0}$, and the average magnification factor, $\mu$, weighted by the surface brightness over the S0(d) image.}
\centering
\begin{tabular}{lccc}
\hline
\hline \\[-0.3em]
Model name & Constraints & $\theta_{\rm E,S0}$ &$\mu$\\ %[+0.5em]
\hline \\[-0.3em]
Esr2$-$3MP (MS) &3 sets + extended arcs  &1.51\arcsec\ & 6.6 \\
                 & of S0, S3                       &        & \\
Esr2$-$3MP (MS$+$Shear) &3 sets + extended arcs  &2.10\arcsec\ & 6.5 \\
 & of S0, S3                       &        & \\
\hline
\end{tabular}
\label{tab:theta_E_s02}
\end{table}

\end{appendix}

\end{document}